\documentclass[11pt,a4paper]{article}
\usepackage{jheppub}
\usepackage[T1]{fontenc}
\usepackage{amsmath}
\usepackage{amssymb}
\usepackage{esint}
\usepackage{amsfonts}
\usepackage{graphicx,color}
\usepackage{slashed}
\usepackage{young}
\usepackage[vcentermath]{youngtab}
\usepackage[utf8]{inputenc}
\usepackage{tensor}
\usepackage{etoolbox}
\usepackage{bbm}
\usepackage{soul}
\usepackage[normalem]{ulem}
\usepackage{cancel}

\makeatletter

\newcommand{\be}{\begin{equation}}
\newcommand{\ee}{\end{equation}}
\newcommand{\bea}{\begin{eqnarray}}
\newcommand{\eea}{\end{eqnarray}}

\renewcommand{\tilde}{\widetilde}
\renewcommand{\hat}{\widehat}

\renewcommand{\d}{\partial}

\newcommand{\NN}{\mathbb{N}}

\def\cL{\mathcal{L}}

\newcommand*\xbar[1]{%
  \hbox{%
    \vbox{%
      \hrule height 0.5pt 
      \kern0.3ex
      \hbox{%
        \kern-0.0em
        \ensuremath{#1}%
        \kern-0.0em
      }%
    }%
  }%
} 

\pdfpageheight\paperheight
\pdfpagewidth\paperwidth

\pdfoutput=1 

    \patchcmd{\maketitle}{\@fpheader}{}{}{}

\title{Asymptotic structure of a massless scalar field and its dual two-form field  at spatial infinity}
\author[a]{Marc Henneaux\footnote{On leave of absence from Coll\`ege de France, Paris},}
\author[b]{and C\'edric Troessaert}
\affiliation[a]{Universit\'e Libre de Bruxelles and International Solvay Institutes, ULB-Campus Plaine CP231, B-1050 Brussels, Belgium}
\affiliation[b]{Max-Planck-Institut f\"{u}r Gravitationsphysik (Albert-Einstein-Institut),
Am M\"{u}hlenberg 1, \\ DE-14476 Potsdam, Germany}

\abstract
{Relativistic field theories with a power law decay in $r^{-k}$ at spatial infinity generically possess an infinite number of conserved quantities because of Lorentz invariance.  Most of these are not related in any obvious way to symmetry transformations of which they would be the Noether charges.  We discuss the issue in the case of a massless scalar field.  By going to the dual formulation in terms of a $2$-form (as was done recently in a null infinity analysis), we relate some of the scalar charges to symmetry transformations acting on the $2$-form and on surface degrees of freedom that must be added at spatial infinity.  These new degrees of freedom are necessary to get a consistent relativistic description in the dual picture, since boosts would otherwise fail to be canonical transformations.  We provide explicit boundary conditions on the $2$-form and its conjugate momentum, which involves parity conditions with a twist, as in the case of electromagnetism and gravity. The symmetry group at spatial infinity is composed of ``improper gauge transformations''.  It is abelian and infinite-dimensional.  We also briefly discuss the realization of the asymptotic symmetries, characterized by a non trivial central extension and point out vacuum degeneracy.}

\makeatother

\begin{document}
\maketitle \flushbottom

\section{Introduction}
\setcounter{equation}{0}

Consider a massless scalar field $\phi$ in flat four dimensional Minkowski space interacting with other fields. Since the scalar field is massless, it is natural to assume that at spatial infinity, it behaves as 
\be
\phi = \frac{\xbar \phi}{r} + \frac{\phi^{(2)}}{r^2} + O(r^{-3})
\label{equ:decay}
\ee
(in 3+1 dimensions), where we use polar coordinates,
\be
ds^2 = - dt^2 + dr^2 + r^2 \xbar \gamma_{AB} dx^A dx^B \, . 
\ee
Here, $ \xbar \gamma_{AB} dx^A dx^B$ is the metric on the round $2$-sphere (in standard ($\theta, \varphi $)-variables, it reads $d \theta^2 + \sin^2 \theta \, d \varphi^2$).  The coefficients in the expansion are allowed to be function of time and of the angles, e.g., $\xbar \phi = \xbar \phi (t, x^A)$.

This behaviour would for instance hold for the Lagrangian
\be
\mathcal{L} = -\frac12 \partial^\mu \phi \partial_\mu \phi -\frac12 \partial^\mu \chi \partial_\mu \chi - \frac12 m^2 \chi^2 + \frac{g}{2} \phi \chi^2  \label{eq:Lagphichi}
\ee
which is the model considered in a similar context in \cite{Campiglia:2017dpg,Campiglia:2017xkp}.  Indeed, the equation for the scalar field is then
\be
\Box \phi + \frac{g}{2} \chi^2 = 0
\ee
For static solutions where the massive field decays exponentially at infinity, the scalar field behaves as in (\ref{equ:decay}).  This is also the behaviour found in \cite{Janis:1968zz,Janis:1970kn} for the coupled Einstein-scalar field equations.  It is therefore natural to adopt the decay (\ref{equ:decay}), but with coefficients that may depend on time for generic configurations.

Now, if the theory is Lorentz invariant, the boundary conditions should be Lorentz invariant.  This means in particular that the above expansion should be preserved under boosts.  This is a non trivial constraint because the boosts blow up linearly in $r$ at infinity. One thus gets, from $\delta_{boosts} \phi = \xi^0 \partial_0 \phi + \xi^m \partial_m \phi$ with $\xi^\mu = O(r)$  and the observation that $\partial_m \phi = O(r^{-2})$, the condition
\be
\partial_0 \xbar \phi = 0
\ee
in order to eliminate the $O(1)$-term in 
$\delta_{boosts} \phi $.  This is an infinite number of conservation laws since $\xbar \phi$ is a function of the angles.

These conservation laws are quite general and merely follow from the decay of the scalar field and Lorentz invariance, which both hold even in the presence of more complicated interactions of the massless scalar field $\phi$.  In that sense, the conservation laws do not give much information on the dynamics.  Nevertherless, it is of interest to understand their significance.

One question that  comes to mind is whether these conservation laws are related to symmetry transformations of the action.  One might (wrongly) think that the answer  is necessarily positive, by arguing that any conserved charge can be expressed in terms of the canonically conjugate variables of the Hamiltonian formalism.  By taking the Poisson bracket of the charges with the canonical variables, one would get a transformation of which the charge would be the Noether charge according to general theorems.    Hence, it would (incorrectly) seem that any conserved quantity could be interpreted as arising from a symmetry transformation.   In order for this reasoning to be correct, however, the conserved quantity must  have well defined Poisson brackets with the canonical variables.  It turns out that a surface term alone does not fulfill this requirement since it does not have well defined functional derivatives by itself.  Hence, in the scalar theory with above Lagrangian, the conserved quantities $\oint_{S^2_\infty}  d^2 x \, \epsilon(x^A) \xbar \phi$ integrated over the $2$-sphere at infinity with an arbitrary smearing function $\epsilon(x^A)$, do not generate any well-defined symmetry. 

In gauge theories, one can sometimes complete the surface term by a bulk term that is proportional to the gauge constraints (and hence does not modify the value of the charge), in such a way that the sum ``bulk term + surface term'' is a well defined generator.  For instance, in electromagnetism,  the infinite number of conserved charges $\oint_{S^2_\infty}  d^2 x \, \epsilon(x^A) \xbar \pi^r$ of the asymptotic electric field $ \bar \pi^r$ over the $2$-sphere at infinity can be completed by bulk terms so that the sum has a well defined action on the canonical variables and generates the infinite dimensional algebra of angle-dependent $u(1)$ transformations \cite{Balachandran:2013wsa,Strominger:2013lka,Barnich:2013sxa,He:2014cra,Henneaux:2018gfi}.  

Not all conserved surface terms can be extended to be well-defined generators in a given formulation of the theory.  To give an example also drawn from the electromagnetic context, the analog magnetic quantities $\oint_{S^2_\infty}  d^2 x \, \eta(x^A) \xbar {\mathcal B}^r$, where $ {\mathcal B}^r$ is the radial magnetic field, which are also conserved, cannot be completed to have well-defined Poisson brackets in the standard electric formulation.  However, by going to the dual, magnetic formulation,  this infinite number of conserved charges can  be completed to have well defined canonical actions, generating a magnetic angle-dependent $u(1)$ \cite{Strominger:2015bla}.  The property of being ``Noether'' depends therefore on the formulation.  

The electromagnetic analysis of  \cite{Henneaux:2018gfi} reveals furthermore that some integration constants become well-defined generators only after the symplectic structure has been modified by a surface term.  This is the case for  the asymptotic radial component $\xbar A_r$ of the vector potential, which is conserved, and which can be interpreted as a symmetry generator provided one introduces further surface degrees of freedom at infinity (without modifying the dynamics of the bulk degrees of freedom) and adds to the symplectic structure a surface term  at infinity.

For the scalar theory, there is no constraint and thus no obvious way to add weakly vanishing terms that would extend the charges $\oint_{S^2_\infty}  d^2 x \, \epsilon(x^A) \xbar \phi$ in the bulk to make them well-defined generators.  The dual formulation, however, involves a $2$-form gauge field and gauge constraints. This suggests exploring the above question in the dual formulation.  This is the objective of this paper.  

We show that scalar charges directly related to the charges exhibited above do have a Noether interpretation in the dual theory.  More precisely, we show that the charges involving the gradient $\partial_A \xbar \phi$  of $\xbar \phi$, i.e., 
$  \oint_{S^2_\infty} \mu^A(x^B) \partial_A \xbar \phi $ where $\mu^A(x^B)$ are arbitrary functions on the $2$-sphere, can be extended in the bulk in such a way that they have well-defined brackets.  That it is the gradients $\partial_A \xbar \phi$ that appear, rather than $ \xbar \phi$ itself, is not surprising given that in the dual theory, the field $\phi$ is not globally defined whenever there are electric sources for the dual $2$-form.  These are strings, and appear as magnetic sources for the scalar field, which is not single-valued as one goes around the strings.  By contrast, the gradients $\partial_i \phi$ are well defined.   We also show that for the interpretation of the scalar charges to be symmetry generators, not only does one need to go to the dual formulation where there are constraints, but one must also introduce surface degrees of freedom at infinity and modify the symplectic form by surface terms.

The study of the scalar charges defined at infinity has been undertaken recently from the point of view of the dual $2$-form theory in \cite{Campiglia:2018see,Francia:2018jtb}\footnote{Earlier investigations of $p$-forms in $2p+2$ dimensions can be found in  \cite{Afshar:2018apx}.}.  These interesting works were motivated by the discovery of the connection between soft theorems and asymptotic symmetries \cite{He:2014laa}, \cite{Campiglia:2017xkp} (for a review and reference to the original literature, see \cite{Strominger:2017zoo}).  They are carried out at null infinity, where these infinite symmetries were first discovered \cite{Bondi:1962px,Sachs:1962wk,Sachs:1962zza} (comprehensive reviews are given in \cite{Madler:2016xju,Alessio:2017lps,Ashtekar:2018lor}).  We consider instead spatial infinity, where complementary aspects of the problem -- including the need for parity conditions to make the symplectic form well defined and their connection with the matching conditions appearing in the null infinity approach -- are interestingly exhibited. 

Among the motivitations for studying the symmetries at spatial infinity, a very strong one comes from the fact that the existence of null infinity with the standardly assumed smoothness properties  is a delicate question in a  spacetime with dynamical metric \cite{Christodoulou:1993uv,Bieri:2009xc,Friedrich:2017cjg,Hintz:2017xxu,Paetz:2018nbd}.  It is then legitimate to wonder whether the infinite-dimensional symmetries exhibited at null infinity would still be present when a sufficiently smooth null infinity does not exist.  The analysis of the dynamics and of the asymptotic symmetries at spatial infinity shows that this is the case and puts therefore the BMS structure on a firm basis independent of the existence of a smooth null infinity \cite{Troessaert:2017jcm,Henneaux:2018hdj,Henneaux:2018gfi,Henneaux:2018cst}.  In particular, the vacuum degeneracy  (non trivial orbit of Minkowski space under the BMS group) clearly appears at spatial infinity without having to invoke gravitational radiation \cite{Henneaux:2018hdj}.

The simplicity of the scalar field equations also serves a pedagogical purpose by shedding direct light on the behaviour of the fields as one goes towards null infinity.  It is indeed easy to explicitly integrate the scalar field equations asymptotically for  given initial data on a spacelike hypersurface.  One finds that even for smooth initial data, the scalar field develops  logarithmic singularities in the null infinity limit.  These can be explicitly computed and related to the behaviour of the initial data under parity.  Similar features are present for electromagnetism and gravity \cite{Henneaux:2018gfi,Henneaux:2018hdj}.

Our paper is organized as follows. In Section \ref{sec:parity}, we study the asymptotic formulation of the massless scalar field.  We point out the need for parity conditions on the leading orders of the field and its conjugate momentum since otherwise, the symplectic structure would have a logarithmic divergence.  We also carefully analyse the behaviour of the scalar field as one goes to null infinity from given initial data on a Cauchy surface and explicitly exhibit the generic non-analytic behaviour in that limit, illustrating the phenomenon discussed in \cite{Christodoulou:1993uv,Bieri:2009xc,Friedrich:2017cjg,Hintz:2017xxu,Paetz:2018nbd}. Just as in the case of electromagnetism and gravity, the natural parity conditions eliminate the leading $\log r$ divergence \cite{Henneaux:2018gfi,Henneaux:2018hdj}.  We then turn in Section \ref{sec:2Form} to the dual $2$-form formulation. We provide boundary conditions, which involve parity conditions with a twist.  This twist is given by an appropriate exterior derivative term.  We then show that because of this twist, Lorentz invariance is problematic unless one modifies the standard symplectic structure by a surface term at infinity.  The most natural way to do so is to introduce also extra surface degrees of freedom at infinity, as in electromagnetism \cite{Henneaux:2018gfi}. The resulting theory possesses an infinite number of asymptotic symmetries (``large'' or ``improper'' gauge transformations) which form an abelian algebra.  We discuss the relation of the corresponding charges with some of the original scalar charges displayed above.  We also compute the canonical realization of the asymptotic symmetry algebra, which we show to be centrally extended. We end up in Section \ref{sec:Conclusions} with conclusions and comments. Three appendices complete the discussion by providing some useful mathematical background (Appendices \ref{app:ultraspherical} and \ref{app:Pseudo}), or discussing some improper gauge fixings (i.e., truncations) of the $2$-form theory (Appendix \ref{app:Minimal}).

\section{Scalar field}
\label{sec:parity}

\subsection{Action in Hamiltonian form -- Boundary conditions}

The Hamiltonian form of the action for the scalar field reads
\be
S [\phi, \pi] = \int dt \left( \int d^3 x  \pi \dot \phi - H \right) \label{eq:Ham0}
\ee
where the Hamiltonian is
\be
H = \int d^3 x \mathcal{H}^\phi, \; \; \; \mathcal{H}^\phi = \frac 12 \left(  \pi^2 + \partial^i \phi \partial_i \phi \right)
\ee
There is also the contribution from the other fields (e.g., the massive field $\chi$ in the above model) but these can be ignored for the present discussion and we shall do so to keep the argument as clear as possible.

We take as boundary conditions  that define phase space the conditions (\ref{equ:decay}) for the scalar field, and $\pi = O(r^{-2})$ for its conjugate ($\pi \sim \dot{\phi}$)
\be
\phi = \frac{\xbar \phi}{r} + \frac{\phi^{(2)}}{r^2} + O(r^{-3}), \; \; \; \pi = \frac{\xbar \pi}{r^2} + \frac{\pi^{(2)}}{r^3} + O(r^{-4}).
\label{equ:decay2}
\ee
We allow off-shell the various coefficients in the expansion in powers of $r^{-1}$ to depend on $t$ and the angles $x^A$.
\subsubsection*{Parity Conditions}
We furthermore impose the parity conditions that  the leading order $\xbar \phi$ should be even under the spatial reflection $x^i \rightarrow - x^i$, and the leading order $\xbar \pi$ should be odd,
\be
\xbar \phi = \rm even, \; \; \; \xbar \pi = \rm odd  \label{eq:ParityCPhi}
\ee
In polar coordinates, the reflection is written $r \rightarrow r$, $x^A \rightarrow - x^A$ (although if the angles are the standard polar angles, one has in fact $\theta \rightarrow \pi - \theta$ and $\varphi \rightarrow \varphi + \pi$). 

These parity conditions make the logarithmic divergence in the kinetic term of the action $\int d^3x \frac{\xbar \pi \, \partial_t \xbar \phi}{r^3}$ actually absent.    [Of course, $\partial_t \xbar \phi$ turns out to vanish so that the potentially divergent integral is zero but we cannot impose $\partial_t \xbar \phi =0$ as a condition on the phase space variables $\phi$ and $\pi$ at a given instant of time.  The definition of phase space should involve only the ``$p$'s and the $q$'s'' and not the ``$\dot{q}$'s''. The equation $\partial_t \xbar \phi =0$ emerges as an equation of motion and holds on-shell, but we do not require it off-shell.]

These parity conditions are the analog for the scalar field of the parity conditions proposed in 
\cite{Regge:1974zd} for gravity and 
\cite{Henneaux:1999ct} for electromagnetism, and generalized in \cite{Henneaux:2018gfi,Henneaux:2018hdj}.

Alternative parity conditions where $\xbar \phi$ would be odd and $\xbar \pi$ would be even would fulfill the same purpose of making the symplectic form finite.  Although they are incompatible with spherical symmetry for $\xbar \phi$, these boundary conditions are mathematically consistent.  It is of interest to consider them also, especially in the study of the behaviour of the fields as one goes towards null infinity.

We finally note that whatever the boundary conditions are, the conserved quantities $\xbar \phi$ do not generate symmetries since they do not have well defined Poisson brackets with the basic canonical variables and hence cannot be viewed as Noether charges.

\subsection{Poincar\'e invariance}

The Poincar\'e transformations acts on the phase space variables as
\be
\delta_{(\xi, \xi^i)} \phi =  \xi \pi + \xi^i \partial_i \phi , \; \; \; 
\delta_{(\xi, \xi^i)} \pi = \partial^i \left( \xi \partial_i \phi \right) + \partial_i \left(\xi^i \pi \right)
\ee
Here, 
\be
\xi \equiv \xi^\perp = b_i x^i + a , \; \; \; \xi^i ={b^i}_j x^j + a^i
\ee
where $b_i$,  $b_{ij} = -b_{ji}$, $a$ and $a^i$ are arbitrary constants. The constants $b_i$ parametrize the Lorentz boosts (the corresponding term $- b^i x^0$ in $\xi^i$ can be absorbed in $a^i$
at any given time), whereas the antisymmetric constants $b_{ij} = -b_{ji}$
parametrize the spatial rotations.  The constants $a$
and $a^i$ parametrize the standard translations.  

The boundary conditions, including the parity conditions, are clearly invariant under the Poincar\'e algebra.

The Poincar\'e transformations are easily verified to leave the symplectic form invariant.  Hence, they are canonical transformations.  This computation uses the parity conditions, since otherwise an unwanted surface term at infinity remains in the variation of the symplectic form (independently of the value of $\partial_t \xbar \phi$).  The Poincar\'e  generators are given by
\be
P_{(\xi, \xi^i)} = \int d^3 x \left[ \xi  \left( \frac12 \pi^2 + \frac12 \partial^i \phi \partial_i \phi \right)  + \xi^i \left(\pi \partial_i \phi\right) \right]
\ee
an expression that is well-defined (converges) thanks again to the parity
conditions.  If the fields did not have definite parity properties, a
logarithmic divergence would appear.  For instance, the first term behaves
as $r^2 dr$ (from the volume element $d^3x$) times $r$ (from $\xi$) times
$r^{-4}$ (from $\pi^2$) $\sim \frac{dr}{r}$, the integral of which
diverges logarithmically, except if the coefficient obtained by integrating over the angles vanishes, which is the case here because $\xbar \pi^2$ is even and $\xi$ is odd.   Parity conditions are thus also needed for finiteness of the Poincar\'e charges.

\subsection{Asymptotic dynamics}
It is easy to write the transformation rules for the asymptotic fields $\xbar \phi$ and $\xbar \pi$.   To that end, it is convenient to go to polar coordinates.  Recalling that the conjugate momentum is a density of weight one, the asymptotic conditions read
\be
\phi = \frac{\xbar \phi}{r} + \frac{\phi^{(2)}}{r^2} + O(r^{-3}), \; \; \; \pi = \xbar \pi + \frac{\pi^{(2)}}{r} + O(r^{-2}).
\label{equ:decay3}
\ee
(where $\xbar \pi \vert_{\rm here} = \sqrt{\xbar \gamma} \, \xbar \pi \vert_{\rm before}$). The Poincar\'e vector fields are
\be
	\xi = rb + a,\quad \xi^r = W, \quad \xi^A = Y^A + \frac 1 r \xbar D^A
	W,  \label{eq:FormOfW}
\ee
with
\be
b = b_1 \sin \theta \cos \varphi + b_2 \sin \theta \sin \varphi + b_3 \cos \theta,
\ee
\be
\hspace{-.5cm} Y = m^1 \left(-\sin \varphi \frac{\partial}{\partial \theta} - \frac{ \cos \theta}{\sin \theta} \cos \varphi \frac{\partial}{\partial \varphi}\right)+ m^2 \left(\cos \varphi \frac{\partial}{\partial \theta} - \frac{ \cos \theta}{\sin \theta} \sin \varphi \frac{\partial}{\partial \varphi}\right) + m^{3} \frac{\partial}{\partial \varphi}
\ee
($b_{ij} = \varepsilon_{ijk} m^k$) and
\be
W = a^1 \sin \theta \cos \varphi + a^2 \sin \theta \sin \varphi + a^3 \cos \theta.
\ee
Here, $\xbar D_A$ is the covariant derivative associated with 
$\xbar \gamma_{AB}$  and $\xbar D^A = \xbar  \gamma^{AB} \xbar D_B$.
The $Y^A$'s are the Killing vectors of the round metric on the unit $2$-sphere, $\cL_Y \xbar
	\gamma_{AB} = 0$.  The function $W$ describes the spatial translations.  One has
\begin{gather}
    \xbar D_A  \xbar D_B b + \xbar \gamma_{AB} b = 0,  \quad
    \xbar D_A \xbar D_B W + \xbar \gamma_{AB} W = 0, \quad \cL_Y \xbar
	\gamma_{AB} = \xbar D_A Y_B + \xbar D_B Y_A = 0.
\end{gather}

The transformation rules of the asymptotic fields under Poincar\'e transformations are
\begin{eqnarray}
&& \delta_{(\xi, \xi^i)} \xbar \phi = b \frac{\xbar \pi}{\sqrt{\xbar \gamma}} + \xi^A \partial_A \xbar \phi, \\
&& \delta_{(\xi, \xi^i)} \xbar \pi = - \sqrt{\xbar \gamma} b \xbar \phi + \partial_A \left( \xbar \gamma^{AB} \sqrt{\xbar \gamma} b \partial_B \xbar \phi \right) + \partial_A \left( \xi^A \xbar \pi\right)
\end{eqnarray}
The asymptotic fields have an autonomous evolution and transform only under boosts and rotations.  They are invariant under translations.   The vacuum configuration $(\xbar \phi = 0, \xbar \pi = 0)$ is invariant and has a trivial orbit.

\subsection{Going to null infinity}
To compare the asymptotic behaviour of the fields at spatial infinity with the asymptotic behaviour of the fields at null infinity, we integrate the equations of motion in hyperbolic coordinates \cite{Ashtekar:1978zz},
\be 
\eta = \sqrt{-t^2 + r^2}, \; \; \; s = \frac{t}{r}
\ee
which cover the region $r > \vert t \vert$.  The inverse transformation reads
\begin{equation}
    t = \eta \frac {s}{\sqrt{1-s^2}}, \quad r = \eta \frac
    {1}{\sqrt{1-s^2}}. \end{equation}
  In hyperbolic coordinates,  the Minkowskian metric reads
\be 
d \eta^2 + \eta^2 h_{ab} dx^a dx^b, \; \; \; \; (x^a) \equiv (s, x^A)
\ee
with 
\be
h_{ab} dx^a dx^b = - \frac{1}{\left(1-s^2\right)^2} ds^2 + \frac{\xbar \gamma_{AB}}{1-s^2} dx^A dx^B
\ee

The equation of motion for $\phi$ is (neglecting the sources, which we can do asymptotically as these are massive)
\begin{equation}
    \d_\mu(\sqrt{-g} g^{\mu\nu} \d_\nu \phi) = \eta \sqrt {-h}
    \Big(\eta^{-1} \d_\eta(\eta^3
    \d_\eta \phi) + \mathcal D^a \mathcal D_a \phi \Big)= 0,
\end{equation}
where $\mathcal D_a$ is the covariant derivative with respect to the metric $h_{ab}$ and $\mathcal D^a = h^{ab} \mathcal D_b$.
The slice $s = 0$ coincides with the Cauchy hyperplane $t=0$, on which $\eta = r$.  We therefore assume that the field has the following asymptotic expansion
\begin{equation}
    \phi(\eta, x^a) = \sum_{k=0} \eta^{-k-1} \phi^{(k)}.
\end{equation}
The homogeneity of the equation of motion implies that each order decouples and fulfills 
\begin{equation}
    \mathcal D^a \mathcal D_a \phi^{(k)} + (k^2-1) \phi^{(k)} = 0,
\end{equation}
which can be rewritten as
\begin{equation}
    -(1-s^2)\d_s^2 \phi^{(k)} + \xbar D^A \xbar D_A \phi^{(k)} + \frac {k^2-1}{1-s^2}
    \phi^{(k)} = 0. \label{eq:ForPhik}
\end{equation}
So, for the free scalar field in hyperbolic coordinates, each order in the expansion in $\eta^{-1}$ fulfills autonomous equations of motion, and not just the leading order.

In order to solve (\ref{eq:ForPhik}), we will develop each of the unknown
functions in spherical harmonics, imposing for the time being no parity condition,
\begin{equation}
    \phi^{(k)} = (1-s^2)^{\frac{1-k}{2}} \sum_{lm} \Theta^{(k)}_{lm}(s)
    Y_{lm}(x^A).
\end{equation}
The parity conditions will be taken care of below. The equation satisfied by the coefficients $\Theta^{(k)}_{lm}$ is then
\begin{equation}
    (1-s^2) \d_s^2 \Theta_{lm}^{(k)} + 2 (k-1) s \d_s \Theta^{(k)}_{lm} +
    \Big[ l(l+1) - k(k-1)\Big] \Theta^{(k)}_{lm} = 0.
\end{equation}
Defining $\lambda = k + \frac 1 2$ and $n=l-k$, we obtain a differential equation that can be straightforwardly  transformed into  the
differential equation for Gegenbauer polynomials, also called ultraspherical polynomials.  These equations are discussed in Appendix \ref{app:ultraspherical} to which we refer the reader for the details, since we shall make a large use of the properties of the solutions of these equations recalled there. 

The zeroth order coefficient $\Theta^{(0)}$ satisfies
Legendre's equation and will be given in terms of Legendre polynomials and
Legendre functions of the second kind.  The general solution for arbirary $k$ is given by
\begin{equation}
    \Theta^{(k)}_{lm} = \Theta_{lm}^{P(k)} \tilde P^{(k + \frac 1
    2)}_{l-k}(s) + \Theta_{lm}^{Q(k)} \tilde Q^{(k + \frac 1
    2)}_{l-k}(s),
\end{equation}
in terms of the $\tilde P^{(k + \frac 1
    2)}_{l-k}(s) $ and $\tilde Q^{(k + \frac 1
    2)}_{l-k}(s)$ of Appendix \ref{app:ultraspherical}, which leads to
\begin{equation}
    \phi = \sum_{k,l,m} \eta^{-k-1}  (1-s^2)^{\frac{1-k}{2}} 
    \Big[\Theta_{lm}^{P(k)} \tilde P^{(k + \frac 1
    2)}_{l-k}(s) + \Theta_{lm}^{Q(k)} \tilde Q^{(k + \frac 1
    2)}_{l-k}(s)\Big]
    Y_{lm}(x^A).
\end{equation}

The easiest way to make contact with null infinity is to introduce the
rescaled radial coordinate $\rho = \eta \sqrt{1-s^2}$ \cite{Fried1,Friedrich:1999wk,Friedrich:1999ax}. The field $\phi$
then takes the form
\begin{equation}
    \phi = (1-s^2) \sum_{k,l,m} \rho^{-1-k} 
    \Big[\Theta_{lm}^{P(k)} \tilde P^{(k + \frac 1
    2)}_{l-k}(s) + \Theta_{lm}^{Q(k)} \tilde Q^{(k + \frac 1
    2)}_{l-k}(s)\Big]
    Y_{lm}(x^A).
\end{equation}
Null infinity is given by the limits $s\to\pm 1$ while keeping $\rho$ and
$x^A$ fixed. As all $\tilde P$ and $\tilde Q$'s are bounded except the
$\tilde Q^{(\frac 1
2)}_n$ that diverge logarithmically, the general expression we obtained
for $\phi$ goes to zero at null infinity.

The link with standard retarded null coordinates $(u, r)$ is given by
\begin{equation}
    s = 1 + \frac u r, \qquad \rho = -2 u - \frac {u^2}{r}, \qquad 1-s^2 =
    -2 u \frac 1 r + O(r^{-2})
\end{equation}
where we take $u<0$ which is relevant to the limit of going to the past of future null infinity. 
Expressing $\phi$ in terms of $u,r$, we get
\begin{multline}
    \phi = \Big(r^{-1} + O(r^{-2})\Big) \sum_{l,m}
    \Theta_{lm}^{Q(0)} \Big(P^{(\frac 1 2)}_l(1) Q^{(\frac 1
    2)}_{0}(1+u/r) + R^{(\frac 1 2)}_l(1)\Big)
    \,Y_{lm}(x^A) +  \frac 1 r \sum_{l,m}
    \Theta_{lm}^{P(0)} P^{(\frac 1
    2)}_{l}(1)
    Y_{lm}(x^A)\\
    +  r^{-1} \sum_{k>0,l,m} (-2u)^{-k} 
    \Big[\Theta_{lm}^{P(k)} \tilde P^{(k + \frac 1
    2)}_{l-k}(1) + \Theta_{lm}^{Q(k)} \tilde Q^{(k + \frac 1
    2)}_{l-k}(1)\Big]
    Y_{lm}(x^A) + O(r^{-2}),
\end{multline}
where we have written the Legendre functions of the second kind $Q^{(\frac 1 2)}_l$
in terms of Legendre polynomials (see Appendix  \ref{app:ultraspherical}). 

Now, one has
\be
    Q_0^{(\frac 1 2)}(s) = \frac 1 2 \log \frac{1+s}{1-s}, 
 \ee
from which one gets 
\begin{eqnarray}
 Q_0^{(\frac 1 2)}(1+u/r) &=& \frac 1 2 \Big(\log r - \log(-u) +
    \log(2 + u/r)\Big) \nonumber \\
    &=& \frac 1 2 \Big(\log( r) + \log 2 - \log(-u)\Big) + 
    o(1).\label{eq:asymptQ12}
\end{eqnarray}

All branches of solutions contribute at most with a $\frac 1 r$
contribution at null infinity except for the leading $Q$ branch
corresponding to leading parity odd solutions. If this branch is non-zero,
the scalar field will have a term of the form $\frac {\log r} r$. It is
interesting to see that this logarithmic branch in $r$ is paired with
a logarithmic divergence in $u$ for the coefficient of the $\frac 1 r$
term. This second linked divergence is coming from the $\log(-u)$ term in
equation \eqref{eq:asymptQ12}.

If the leading $Q$ branch is absent, i.e. if $\Theta_{lm}^{Q(0)} = 0$,
then at null infinity, we have $\phi = \xbar \phi(u,x^A) \frac 1 r +
O(r^{-2})$ and, in the limit $u\to -\infty$, the leading term tends to a
function on the circle controlled by the leading $P$ branch
\begin{equation}
    \lim_{u\to -\infty} \xbar \phi(u, x^A) = \sum_{l,m}
    \Theta_{lm}^{P(0)} Y_{lm}(x^A).
\end{equation}
This is the asymptotic behaviour near null infinity assumed in \cite{Campiglia:2018see,Francia:2018jtb}.

\subsubsection*{Parity conditions and matching conditions}

The asymptotic behaviour on spacelike hyperplanes of the explicit solution found above in
hyperbolic coordinates can easily be worked out by considering the hyperplane $s=0$.  One finds 
 \begin{flalign}
    \xbar \phi & = \lim_{r\to\infty} r \phi = \sum_{l,m} 
    \Big[\Theta_{lm}^{P(0)} P^{(\frac 1
    2)}_{l}(0) + \Theta_{lm}^{Q(0)} Q^{(\frac 1
    2)}_{l}(0)\Big]
    Y_{lm}(x^A),\\
    \xbar \pi & = \lim_{r\to\infty} r^2 \sqrt{ \xbar \gamma}\, \d_t\phi = \sin \theta \sum_{l,m} 
    \Big[\Theta_{lm}^{P(0)} \d_s P^{(\frac 1
    2)}_{l}(0) + \Theta_{lm}^{Q(0)} \d_s Q^{(\frac 1
    2)}_{l}(0)\Big]
    Y_{lm}(x^A),
\end{flalign}
reproducing (\ref{equ:decay3}).

Due to the parity properties
 $$Y_{lm}(-x^A) = (-1)^l Y_{lm}(x^A )$$
 ($ \Leftrightarrow Y_{lm}(\pi- \theta, \varphi + \pi) = (-1)^l Y_{lm}( \theta, \varphi )$)
  of the spherical harmonics, and  
  $$P^{(\frac 1
    2)}_{l}(-s) = (-1)^l P^{(\frac 1
    2)}_{l}(s), \qquad Q^{(\frac 1
    2)}_{l}(-s) = (-1)^{l+1} Q^{(\frac 1
    2)}_{l}(s),$$
  of the Legendre polynomials/functions,  which imply 
   $$P^{(\frac 1
    2)}_{l}(0) =0, \qquad \partial_s Q^{(\frac 1
    2)}_{l}(0) =0 \quad \textrm{for odd $l$'s},$$ and 
$$Q^{(\frac 1
    2)}_{l}(0) =0, \qquad \partial_s P^{(\frac 1
    2)}_{l}(0) =0 \quad \textrm{for even $l$'s},$$    
  we see that
$\Theta^{P(0)}_{lm}$ control the even part of $\xbar \phi$ and the odd
part of $\xbar \pi$ while $\Theta^{Q(0)}_{lm}$ control the other parity components. 

Thus, in order to fulfill (\ref{eq:ParityCPhi}), we must take $\Theta^{Q(0)}_{lm} = 0$, i.e., set 
the leading $Q$ branch equal to zero.   This eliminates, as we have seen, the dominant
logarithmic behaviour at null infinity. This is in line with the investigations performed at null infinity, where this singular behaviour is usually assumed
to be absent. As in the cases of electromagnetism and gravity \cite{Henneaux:2018gfi,Henneaux:2018hdj}, the parity conditions at spacelike infinity eliminate the leading singularities at null infinity.

Furthermore, the solutions $\phi^{(0)}(s, x^A)$ with $\Theta^{Q(0)}_{lm} = 0$ fulfill $\phi^{(0)}(s, x^A) = \phi^{(0)}(-s, -x^A)$ due to the combined parity properties of the spherical harmonics and the Legendre polynomials.  This implies the matching conditions  
\be
\lim_{u\to -\infty} \xbar \phi(u, x^A) =  \lim_{v\to \infty} \xbar \phi(v, -x^A)
\ee
 of \cite{Strominger:2017zoo} relating the leading order of the fields on the past boundary of future null infinity with the the leading order of the fields on the future boundary of past null infinity at the antipodal points.  This is again just as in the cases of electromagnetism and gravity \cite{Henneaux:2018gfi,Henneaux:2018hdj}.
 
 We should stress in closing this subsection that even though the leading divergence is eliminated at null infinity by the condition $\Theta^{Q(0)}_{lm} = 0$, subleading terms of the form $\frac{\log r}{r^m}$ ($m\geq 2$) will generically appear.  In many analyses of the asymptotic properties at null infinity, such terms are excluded, a condition that is too strong and unnecessary from the point of view of the description of the dynamics on Cauchy hypersurfaces.   It is interesting to point out in this respect that the alternative parity conditions obtained by setting the leading $P$ branch equal to zero and keeping the leading $Q$ branch are perfectly regular from the point of view of spacelike infinity and leads therefore also to a consistent, self-contained Hamiltonian description (it is only at null infinity that logarithmic divergences appear).

\section{Two-form gauge field}
\label{sec:2Form}
We now turn to the dual formulation.

In four dimensions, a scalar field is dual to a $2$-form gauge field with action
 \be
 S[B_{\mu \nu}] = -\frac16 \int d^4 x \, C_{\lambda \mu \nu} C^{\lambda \mu \nu} 
 \ee
 Here, the curvature $C_{\lambda \mu \nu}$ is defined by
 \be
 C_{\lambda \mu \nu} = \partial_\lambda B_{\mu \nu} + \partial_\mu B_{\nu \lambda} + \partial_\nu B_{\lambda \mu}
 \ee
 and is invariant under the gauge transformations 
 \be
 \delta B_{\mu \nu} = \partial_\mu \epsilon_\nu - \partial_\nu \epsilon_\mu
 \ee
 which are reducible, since 
 \be
 \epsilon_\mu = \partial_\mu \eta
 \ee
 yields $\delta B_{\mu \nu} = 0$.

\subsection{Hamiltonian and Constraints}

The action in Hamiltonian form reads
\be
S[B_{ij}, \pi^{ij}, B_{0i}] = \int d^4 x \left\{\pi^{ij} \partial_t B_{ij} - B_{0i} {\mathcal G}^i - \left( \frac12 \pi^{ij} \pi_{ij} + \frac16 C_{ijk} C^{ijk} \right)\right \} \label{eq:Ham2}
\ee
where $\pi^{ij}$ are the momenta conjugate to $B_{ij}$.  The $B_{0i}$ are the Lagrange multipliers for the constraints ${\mathcal G}^i \approx 0$, with
\be
{\mathcal G}^i = -2 \partial_j \pi^{ji}
\ee
(``Gauss law'')

One can relate the $2$-form action (\ref{eq:Ham2}) to the scalar field action (\ref{eq:Ham0}) by the following change of variables:
\be 
\pi^{ij} = \frac{1}{\sqrt{2}}\varepsilon^{ijk} \partial_k \phi, \; \; \; C_{ijk} = \frac{1}{\sqrt{2}}\varepsilon_{ijk} \pi  \label{eq:CofV}
\ee
As shown in Appendix \ref{app:Minimal}, the two actions differ by a surface term at the time boundaries and a surface term at spatial infinity.  The suface term at the time boundaries depends on the arguments of the transition amplitude (what is kept fixed at the time boundaries in the path integral - or in the action principle) and must be determined on this ground.  The surface term at spatial infinity is crucial for a proper definition of the symplectic structure. 

The change of variables (\ref{eq:CofV}) needs some qualifications.  The field $\phi$ is well-defined because the conjugate momenta $\pi^{ij}$ fulfill the constraints ${\mathcal G}^i \approx 0$, which are the necessary   conditions for $\partial_k \phi = \frac{1}{\sqrt{2}}\varepsilon_{kij} \pi^{ij}$ to be integrable.  If there were electric sources for the $2$-form (strings), which are magnetic sources for $\phi$, these would modify Gauss law and the integrability conditions for $\partial_k \phi$ would fail at the location of the sources.  Accordingly, the integral of $\partial_k \phi$ on a closed line linking the source would not vanish and $\phi$ would not return to its original value for such a loop.    So, while the gradients of the field $\phi$ are well-defined, its zero-mode is multiple-valued. One way to deal with this problem is to proceed ``\`a la Dirac'' \cite{Dirac:1948um} and introduce ``Dirac membranes'' attached to the strings \cite{Teitelboim:1985ya,Teitelboim:1985yc}.  The Dirac membranes are pure gauge objects and are only needed in the scalar field formulation.

By contrast, the electric sources for $\phi$, which are magnetic sources for the $2$-form, do not lead to a difficulty in the change of variables (\ref{eq:CofV}). For any given $\pi$, the equations (\ref{eq:CofV}) admit a solution for $B_{ij}$.  What happens is that only the definition of the momenta conjugate to $B_{ij}$ is affected by the presence of the sources for $\phi$.  But the relationship $\pi^{ij} \leftrightarrow \dot{B}_{ij}$ is not used in (\ref{eq:CofV}), which is therefore unchanged.  The relationship $\pi^{ij} \leftrightarrow \dot{B}_{ij}$ emerges as the Hamiltonian equation of motion for $\pi^{ij}$. For instance, in the case of the Lagrangian (\ref{eq:Lagphichi}), the coupling term $\int d^3x \phi \chi^2$ in the Hamiltonian for the scalar field leads to the non local term $ \int d^3x \triangle^{-1}\left(\varepsilon_{ijk} \partial^k \pi^{ij} \right) \chi^2$ in the Hamiltonian for the $2$-form gauge field, which modifies the conjugate momentum $\pi^{ij}$ by the non-local contribution $\sim \triangle^{-1}\left(\varepsilon^{ijk} \partial_k (\chi^2) \right)$.

\subsubsection*{Asymptotic conditions -- Parity conditions}

The fall-off of the scalar field and its conjugate momentum implies (i) that up to an exterior derivative, the $2$-form components $B_{ij}$ decay in Cartesian coordinates as $\frac1r$ with a leading term that is even under parity\footnote{This means that $\xbar B$ is an odd pseudo-$2$-form and that its conjugate is an even pseudo-bivector (density), see Appendix \ref{app:Pseudo}.}; and (ii) that the components $\pi^{ij}$ decays in Cartesian coordinates as $\frac{1}{r^2}$ with a leading term that is odd under parity.

The  exterior derivative term $\partial_i \Lambda_j - \partial_i \Lambda_j $ is admissible  in the asymptotic fall-off of $B_{ij}$ since it drops out from $C_{ijk}$.  One could a priori think that $\Lambda_i$ is completely arbitrary.  There are constraints, however,  coming from the fact that the symplectic form and the charges should be finite, which are fulfilled if $\Lambda_i = O(r^0)$ as we shall assume here (although we have not explored whether a more flexible asymptotic behaviour could be consistently considered).  This leads us to the following boundary conditions at spatial infinity, which we express in polar coordinates\footnote{Parity properties of pseudo-$2$-forms are as follows.  If the leading components $\xbar B_{ij}$ in Cartesian coordinates are even (odd pseudo-$2$-form), then the leading components in the coordinate system $(r, x^A)$ with $r \rightarrow r$ and $x^A \rightarrow - x^A$ (e.g., $(x^A) = (x,y)$) have ``exchanged'' parity properties because the frame $\{\d_r, \d_A\}$ has same orientation as its image $\{\d_r, -\d_A\}$ under parity (while $\{\partial_i\}$ and $\{-\partial_i\}$ have opposite orientation).  Thus, $\xbar B_{rA}$ will be even up to an exterior derivative while $\xbar B_{AB}$ will be odd. Similarly, $\xbar \pi^ {rA}$ will be odd while  $\xbar \pi^{AB}$ will be even. Parity properties of components in spherical coordinates $(r, \theta, \varphi)$ will be however the standard ones.  See Appendix \ref{app:Pseudo} for more information.},
\begin{eqnarray}
    && B_{rA} = \xbar B_{rA} + O\left(\frac{1}{r}\right), \quad 
    B_{AB} = r \xbar B_{AB} + O(1), \label{eq:falloff1}\\
    && \pi^{rA} = \frac{\xbar \pi^{rA}}{r}  + O\left(\frac{1}{r^2}\right), \quad
    \pi^{AB} = \frac{\xbar \pi^{AB}}{ r^2}  + O\left(\frac{1}{r^3}\right) \label{eq:falloff2}
\end{eqnarray}
with 
\begin{eqnarray}
&&\xbar B_{rA} = \xbar B^{even}_{rA} + \xbar \Lambda_A - \partial_A \xbar \Lambda_r ,\label{eq:Parity1}\\
&&\xbar B_{AB} =  \xbar B_{AB}^{odd} + \partial_A \xbar \Lambda_B - \partial_B \xbar \Lambda_A , \label{eq:Parity2}\\
&&\xbar \pi^ {rA}  =  \xbar \pi^{rA}_{odd} , \quad  \xbar \pi^{AB} =   \xbar \pi^{AB}_{even} \label{eq:Parity3}
\end{eqnarray}
where
\begin{eqnarray}
&&\xbar B^{even}_{rA} (-x^A) =  \xbar B^{even}_{rA} (x^A), \\
&&\xbar B^{odd}_{AB} (-x^A) =  -\xbar B^{odd}_{AB} (x^A) ,\\
&&\xbar \pi^ {rA}_{odd} (-x^A) =  -\xbar \pi^{rA}_{odd} (x^A), \quad  \xbar \pi^{AB}_{even} (-x^A) =   \xbar \pi^{AB}_{even} (x^A).\label{eq:Parity26}
\end{eqnarray}
Here
\be
\Lambda_r = \xbar \Lambda_r + O(\frac{1}{r}), \qquad \Lambda_A = r \xbar \Lambda_A + O(1)
\ee
All fields with an overbar depend again only on the angles and time, i.e., are time-dependent fields on the sphere at infinity.  Since the even part of $\xbar \Lambda_A$ and odd part of $\xbar \Lambda_r$ can be absorbed through redefinitions of $\xbar B_{ij}$, one could assume that $\xbar \Lambda_A$ is odd and $\xbar \Lambda_r$ is even, but we shall not do so.

In addition to the asymptotic fall-off (\ref{eq:falloff1}), (\ref{eq:falloff2}) and the parity conditions (\ref{eq:Parity1})-(\ref{eq:Parity26}), we also impose that the constraint vector-densities $\mathcal G^i \equiv \partial_j \pi^{ij} $ decay one order faster than the decay that follows from (\ref{eq:falloff2}), i.e.,  $\mathcal G^r = o(r^{-1})$ and $  \mathcal G^A = o(r^{-2}) $, which is equivalent to 
\begin{equation}
    \d_A \xbar \pi^{Ar} = 0, \quad \xbar\pi^{rA} - \d_B\xbar\pi^{BA} = 0.
\end{equation}

The parity conditions supplemented by these constraint conditions make the symplectic form finite.  They are the analogs of the parity conditions for electromagnetism and gravity  considered in  \cite{Henneaux:2018gfi,Henneaux:2018hdj}.
As for the scalar field, opposite parity conditions can be consistently defined (which would also cover the case of a true (non ``pseudo'') $2$-form) but we shall not develop them explicitly here, referring simply to the scalar field analysis.  

\subsection{Boosts  -- Surface degrees of freedom}

\subsubsection*{Problem with boosts}
The Poincar\'e transformation laws are given by
\begin{equation}
    \delta_\xi B_{ij} = \frac \xi {\sqrt g} \pi_{ij} + \xi^m C_{mij}
      + \d_i \zeta_j - \d_j \zeta_i, \qquad
    \delta_\xi \pi^{ij} = \d_k \left( \xi \sqrt g \, C^{kij}\right) + 3
    \d_k(\xi^{[k}\pi^{ij]}),
\end{equation}
where $\zeta_r = \xbar \zeta_r + O(r^{-1})$ and $\zeta_A = r \xbar \zeta_A
+ O(1)$ is the parameter of a gauge transformation which one can include in the definition of the Poincar\'e transformations of the gauge-variant fields $B_{ij}$.  Definite choices will be made below.  It is easy to verify that the asymptotic conditions are Poincar\'e invariant.

The Poincar\'e transformations are also easily verified to leave the symplectic form $\sigma^{\textrm{bulk}} \equiv \int d^3 x \, d_V \pi^{ij} \wedge d_V B_{ij}$ invariant, except the boosts, which present subtleties.  We thus focus on boosts from now on, for which $\xi^i = 0$ and $\xi = b r$.
Asymptotically, these read explicitly
\begin{gather}
    \delta_\xi \xbar B_{rA} = \frac b {\sqrt {\xbar \gamma}} \xbar \gamma_{AB} \xbar \pi^{rB}
      + \xbar \zeta_A - \d_A \xbar \zeta_r, \quad
    \delta_\xi \xbar B_{AB} = \frac b {\sqrt {\xbar \gamma}} \xbar \pi_{AB} 
      + \d_A \xbar \zeta_B - \d_B \xbar \zeta_A, \\
    \delta_\xi \xbar \pi^{rA} = \sqrt {\xbar \gamma}\,  \xbar \gamma^{AC} \, \xbar D^B \left( b \,
    \xbar C_{rCB}\right) , \\
    \delta_\xi \xbar \pi^{AB} = - b \sqrt{\xbar \gamma} \, \xbar \gamma^{AC} \xbar \gamma^{BD}
   \xbar C_{rCD}.
\end{gather}
where
\be
\xbar C_{rCD} = \xbar B _{CD} - \partial_C \xbar B _{rD} + \partial_D \xbar B_{rC}.
\ee

We note incidentally that these transformations present a striking difference with respect to the corresponding ones for gravity and electromagnetism.  They mix radial and angular components of the dynamical variables.  For that reason, ``twisted parity conditions'' where one modifies the parity of the angular components without changing the parity of the radial ones \cite{Henneaux:2018cst}, do not appear to be available.

One finds for the variation of the symplectic form under boosts,
\begin{eqnarray}
\delta_\xi \Omega^{\textrm{bulk}} &=& d_V \left( i_\xi \Omega^{\textrm{bulk}}\right) \nonumber \\
&=& \oint d^2x \, \sqrt{\xbar \gamma} \,  b \, \xbar \gamma^{CA} \, \xbar \gamma^{DB} \, d_V \xbar C_{rCD} \,  d_V \xbar B _{AB} \label{eq:VarSigma}
\end{eqnarray}
where we have omitted the $\wedge$ symbol and where $d_V$ is the exterior derivative in field space. The function $\xbar C_{rCD}$ is odd, as is $b$.  This means that only the even part of $\xbar B _{AB}$ contributes to the integral.  This even part is not zero, however, and although it is given by an exterior derivative,  the integral (\ref{eq:VarSigma}) does not vanish.  Something must therefore be done for the boosts to be canonical transformations.  

Various possibilities exist.  A minimal one, which does not require extra variables, is given in Appendix \ref{app:Minimal}.  We explain here a different route, which is in the line of what one does for electromagnetism (for which the minimal version does not exist) \cite{Henneaux:2018gfi}.   This approach is richer, in the sense that it displays more symmetries and enables one to view some of the conserved quantities exhibited above as corresponding Noether charges.  [In fact, the minimal approach can be viewed as resulting from an improper -- and hence non permissible -- gauge fixing of the non-minimal one, which eliminates physical degrees of freedom - see Appendix \ref{app:Minimal}.]

In this non minimal approach, one introduces surface degrees of freedom at spatial infinity with appropriate Lorentz transformations  and adds  surface terms to the bulk symplectic form  in such a way that the total symplectic form is invariant.  

There are two equivalent technical ways to introduce the surface degrees of freedom at infinity. One can either just introduce these degrees of freedom only at spatial infinity, or one can introduce fields in the bulk that match these degrees of freedom  at infinity, with the condition that their conjugate momentum is contrained to vanish.  In this manner, there is no new physical bulk degree of freedom that is introduced (the new bulk degrees of freedom are pure gauge).   But with the surface modification of the symplectic form,  some non trivial degrees of freedom  remain at infinity.  Both methods are described for electromagnetism in \cite{Henneaux:2018gfi} and shown to be equivalent.  We shall follow below the second procedure. 

\subsubsection*{New surface degrees of freedom}

Inspired by the electromagnetic results, we add to the original $2$-form canonical pair $(B_{ij}, \pi^{ij})$ subject to  Gauss' law $- 2 \partial_i \pi^{ij} \approx 0$, the following conjugate pairs,
\begin{equation}
    (p^{i}, \Psi_i) \, \qquad
    (\pi_\Phi, \Phi) 
\end{equation}
with the constraints 
\be
p^{i} \approx 0 , \qquad \pi_\Phi \approx 0.
\ee
The first pair is the direct analog of the pair $(\Psi, \pi_\psi)$ introduced in  \cite{Henneaux:2018gfi}  for electromagnetism. The second pair arises because of the reducibility of the gauge transformations.

In addition to the above asymptotic fall-off of the pair $(B_{ij}, \pi^{ij})$, we impose the asymptotic behaviour
\begin{gather}
    \Psi_r =  \frac{\xbar \Psi_r}{r} + O\left(\frac{1}{r^2}\right), \quad
    \Psi_A = \xbar \Psi_A + O\left(\frac{1}{r}\right), \quad 
    p^r = O\left(\frac{1}{r}\right), \quad p^A = O\left(\frac{1}{r^2}\right), \nonumber\\
    \Phi =  \frac{\xbar \Phi}{r} + O\left(\frac{1}{r^2}\right), \quad \pi_\Phi = O\left(\frac{1}{r}\right).
\end{gather}
 (in polar coordinates) on the new fields.

\subsubsection*{Symplectic structure}
The complete symplectic structure is taken to be
\begin{multline}
    \Omega = \int d^3x \left( d_V \pi^{ij} d_V B_{ij} + d_V p^i d_V
    \Psi_i + d_V \pi_\Phi d_V \Phi\right)\\ +
    \oint d^2x \sqrt{\xbar \gamma}\left( -2 \xbar \gamma^{AB} d_V \xbar B_{rA} d_V \xbar
    \Psi_B +2 d_V \xbar \Psi_r d_V \xbar \Phi \right).
\end{multline}
and differs from the above $\Omega^{\textrm{bulk}}$ by terms that vanish on the constraint surface and by boundary terms.  This is the standard ``dp  dq'' symplectic form, modified by surface terms that are such that the boosts are canonical transformations (see below). 

It is not necessary to introduce parity conditions on the new variables to make the symplectic form finite because the new momenta decrease sufficiently fast at infinity.   

\subsubsection*{Hamiltonian and action}

The Hamiltonian $H$  is taken to be
\be
H = \int d^3 x  \mathcal H
\ee
with
\be
 \mathcal H  = \frac 1 {2\sqrt g} \pi^{ij} \pi_{ij} 
      + \frac {\sqrt g} 6 C_{ijk} C^{ijk} - B_{ij} \nabla^i p^j 
      + \d_i \Phi p^i + \nabla^i \Psi_i \pi_\Phi - 2 \Psi_i \d_j \pi^{ji},
      \label{eq:HamDensity0}
\ee 
It is a direct generalization of the Hamiltonian taken in the electromagnetic case.  The integrand weakly coincides with the energy density
$$
\frac 1 {2\sqrt g} \pi^{ij} \pi_{ij} 
      + \frac {\sqrt g} 6 C_{ijk} C^{ijk}
$$
as it should.  All the extra terms vanish with the constraints.  Their specific form has been chosen for later convenience.

The action reads
\begin{multline}
S[B_{ij}, \pi^{ij}, \Psi_i, p^i , \Phi, \pi_\Phi; \rho^a] = \int dt \left[ \int d^3x \left( \pi^{ij} \partial_t B_{ij} +  p^i \partial_t
    \Psi_i +  \pi_\Phi \partial_t \Phi\right) - H - \int d^3 x \, \rho^a \, G_a\right]\\ +
   \int dt  \oint d^2x \sqrt{\xbar \gamma}\left( -2 \xbar \gamma^{AB}  \xbar B_{rA} \partial_t \xbar
    \Psi_B +2  \xbar \Psi_r \partial_t  \xbar \Phi \right) \label{eq:FullAction2B}
\end{multline}
where $\rho^a$ stands for all the Lagrange multipliers enforcing the constraints collectively denoted $G_a \approx 0$.

The new fields boundary fields fulfill 
\be 
\partial_t \xbar \Psi_B = 0, \qquad \partial_t \xbar \Psi_r = 0, \qquad \partial_t  \xbar \Phi = 0
 \ee
and one can easily check that they affect the equations of motion of the original $2$-form field only by terms that vanish when the constraints are taken into account.  Note in particular that the equation $\partial_t \xbar B_{rA} =0$, which follows from extremization of the action with respect to $\xbar  \Psi_A$, is a consequence of the equations for the $2$-form. 

As in the case of electromagnetism, $\Psi_i$ and the temporal components of the $2$-form both multiply Gauss' law in the action.  There is therefore some redundancy, which can be eliminated by identifying $\Psi_i $ and $B_{0i}$ and keeping only one term proportional to Gauss' constraint in the action\footnote{Technically, one might describe the procedure as follows.  Following strictly Dirac's method for the original second-order $2$-form action, one would introduce a conjugate momentum for all $2$-form components including $B_{0i}$. There is then a primary constraint $\pi_{B_{0i}} \approx 0$. The ``total Hamiltonian''  and the ``total action'' involves only the primary constraints multiplied by arbitrary Lagrange multipliers.  When setting $B_{0i} = \Psi_i$ (and $\pi_{B_{0i}}  = p_i$) and keeping only $\Psi_i \mathcal{G}^i$  (without the additional  copy $B_{0i} \mathcal{G}^i$), one effectively sticks to this total action (with the extra ``non minimal'' variables $(\Phi, \pi_\phi)$ added because of reducibility).  One does not go to the (physically equivalent) extended formalism with secondary constraints multiplied also by arbitrary Lagrange multipliers, which exhibits more explicitly all the gauge freedom.  The condition $B_{0i} = \Psi_i$ is a gauge condition that reduces the ``extended formalism'' to the ``total formalism''.}.   With that identification, the surface degrees of freedom  $\xbar \Psi_i$ are the $O(r^{-1})$ pieces of $B_{0i}$. As we shall see, they are not pure gauge.

In a gauge system, there is always some ambiguity in the Hamiltonian, to which one can always add combinations of the constraints, corresponding to the fact that a time translation can be accompanied by gauge transformations.  The particular choice of Hamiltonian  (\ref{eq:HamDensity0}) was guided by the identification $\Psi_i = B_{0i}$ and the request to implement the (generalized) Lorentz gauge  $\partial_\mu B^{\mu \nu} + \partial^\nu \Phi=0$.  This request makes Lorentz invariance easy to control (see below).  The Lorentz gauge conditions determine the evolution of the new fields, since they read in $3+1$ notations $\d_t \Psi_i = \partial^j B_{ji} + \partial_i \Phi$ and $\d_t \Phi = \partial^i \Psi_i$. These  are indeed just the equations of motion following from our choice of Hamiltonian, as announced  (in other words, the Lorentz gauge is equivalent to $\rho^a = 0$ for the Lagrange multipliers associated with the new constraints once the identification of $\Psi_i$ with $B_{0i}$ is made).  The extra term $\partial_\nu \Phi$ is introduced in the (generalized) Lorentz gauge to avoid the constraint $\partial^i \Psi_i = 0$ that would follow from the condition $\partial_\mu B^{\mu \nu} = 0$ without $\Phi$-term.  We  want the Lorentz gauge conditions to be only evolution equations, as in electromagnetism. The need for $\Phi$ is a feature that is present because the gauge symmetries of the $2$-form are redundant.  It has no analog in electromagnetism.  Note that the Lorentz gauge implies that the leading orders $\xbar \Psi_i$ and $\xbar \Phi$ are indeed constant, i.e., $\d_t \xbar \Psi_i = 0$ and $\d_t \xbar \Phi = 0$ are the first terms in the expansion of the Lorentz gauge conditions.

\subsubsection*{Boosts}
We now extend the boost transformation laws to include the new fields.  These transformations must fulfill the following requirements:
\begin{itemize}
\item They must be canonical transformations, i.e., $\mathcal{L}_\xi \Omega = d_V (i_\xi \Omega) = 0$.
\item They must reduce to the previous transformations for the $2$-form fields and their momenta, at least when the constraints hold.
\item They must preserve the constraint surface.
\end{itemize}
The following transformations fulfill these requirements,
\begin{gather}
    \label{eq:fullponcareI}
    \delta_\xi B_{ij} = \frac \xi {\sqrt g} \pi_{ij} 
      + \d_i (\xi\Psi_j) - \d_j (\xi \Psi_i), \qquad
    \delta_\xi \pi^{ij} = \d_k \left( \xi \sqrt g \, C^{kij}\right)+ \xi
    \nabla^{[i} p^{j]},\\
    \delta_\xi \Psi_i = \nabla^j( \xi B_{ji}) + \xi \d_i \Phi, \qquad
    \delta_\xi p^i = 2 \xi \d_j \pi^{ji} + \nabla^i(\xi \pi_\Phi), \\
    \delta_\xi \Phi = \xi \nabla^i \Psi_i, \qquad
    \delta_\xi \pi_\Phi = \d_i(\xi p^i),
\end{gather}
which implies
\begin{gather}
    \delta_\xi \xbar B_{rA} = \frac b {\sqrt {\xbar \gamma}} \xbar \gamma_{AB} \xbar \pi^{rB}
      + b \xbar \Psi_A - \d_A(b \xbar \Psi_r), \\
    \delta_\xi \xbar B_{AB} = \frac b {\sqrt {\xbar \gamma}} \xbar \pi_{AB} 
      + \d_A (b\xbar \Psi_B) - \d_B (b\xbar \Psi_A), \\
    \delta_\xi \xbar \pi^{rA} = \sqrt {\xbar \gamma}\,  \xbar \gamma^{AC} \, \xbar D^B \left( b \,
    \xbar C_{rCB}\right) , \\
    \delta_\xi \xbar \pi^{AB} = - b \sqrt{\xbar \gamma} \, \xbar \gamma^{AC} \xbar \gamma^{BD}
   \xbar C_{rCD}, \\
    \delta_\xi \xbar \Psi_r = - \xbar D^A( b\xbar B_{rA}) - b \xbar \Phi, \qquad
    \delta_\xi \xbar \Psi_A = \xbar D^B (b \xbar B_{BA}) + b \xbar B_{rA} +
    b \d_A \xbar \Phi,\\
    \delta_\xi \xbar \Phi = b(\xbar \Psi_r + \xbar D^A \xbar \Psi_A).
\end{gather}

[The above transformations were obtained by demanding that they should coincide with the Lie derivatives of $B_{\mu \nu}$ and $\Phi$ (which is a scalar) under boosts, if one uses the equations of motion of the fields to eliminate their time derivatives.]

One has $d_V (i_\xi \Omega) = 0$ so that $i_\xi \Omega = d_V P_{\xi,0}$ with 
a generator $P_{\xi,0}$ given by
\begin{flalign}
    P_{\xi,0} & = \int d^3x \, \xi \mathcal H + \mathcal B_{\xi,0}, \label{eq:GenBoosts}\\
    \mathcal B_{\xi,0} & = \oint d^2x \, 2 b \xbar \pi^{rA} \xbar \Psi_A 
      + \oint d^2x \sqrt \gamma \, b \Big( \xbar \Psi_A \xbar \Psi^A
      - \xbar B_{rA} \xbar B^{\phantom r A}_r + 2 \xbar \Psi_r \xbar D^A \xbar \Psi_A 
    \nonumber \\ & \qquad
      + (\xbar \Psi_r)^2 + (\xbar\Phi)^2 - 2 \xbar B_{rA} \xbar D^A \xbar \Phi 
      - \frac 1 2 \xbar B_{AB} \xbar B^{AB} 
      + \xbar B^{AB} (\d_A \xbar B_{rB} - \d_B \xbar B_{rA})\Big).
\end{flalign}

\subsubsection*{Poincar\'e generators}
The other Poincar\'e transformations have also a well defined generator.  As is well known, transformations of the fields under a symmetry are defined up to a gauge transformation in any gauge theory.  For spatial translations and rotations, we adjust the gauge transformation in such a way that the action of these spatial symmetries on the fields is the ordinary Lie derivative, 
 where $\Psi_i$ is a spatial vector and $\Phi$ is a spatial scalar.   Spatial translations and rotations are then generated by
\begin{flalign}
	\label{eq:2FoRotGen}
	P_{0,\xi^i} & = \int d^3x( \pi^{ij} \cL_{\xi^k} B_{ij}+ p^i \cL_{\xi^k} \Psi_i + \pi_\phi \cL_{\xi^k} \Phi) \\
	&-2 \oint d^2x \sqrt{\xbar\gamma} \Big(\xbar B_{rA} \xbar \gamma^{AB} (Y^C \partial_C \xbar \Psi_B + \d_B Y^C \xbar \Psi_C)+ \xbar \Phi Y^A\d_A\xbar \Psi_r\Big).
\end{flalign}
The boundary term in the expression of the rotation charges (angular momentum) is essential  in order to fulfill $i_{\xi^k} \Omega = d_V P_{0,\xi^k}$.

For the generator of time translations, we take the same expression (\ref{eq:GenBoosts}) as for the boosts, but with $\xi = a$.  The boundary term is then zero and the generator reads
\be
    \label{eq:fullponcareEnd}
	P_{a,0}= a\int d^3x\, \mathcal{H}.
\ee
This amounts again to a specific choice of the improper gauge transformation included in what is meant by a ``time translation'' and is again a matter of choice.  Our choice leads to the simple algebra (\ref{eq:2FoAlgebraI}), (\ref{eq:2FoAlgebraII}).  The generator (\ref{eq:GenBoosts}) is thus generally valid for
\begin{equation}
	\xi = b(x^A)r + a.
\end{equation}

The algebra of the Poincar\'e generators is given by
\begin{gather}
	\label{eq:2FoAlgebraI}
	\{P_{\xi_1, \xi^i_1},P_{\xi_2, \xi^i_2}\} = P_{\hat\xi, \hat
	\xi^i}, \\
	\hat \xi = \xi_1^i \d_i \xi_2 - \xi_2^i \d_i \xi_1,
	\label{eq:2FoAlgebraII}
	\quad \hat \xi^i = \xi_1^j \d_j \xi_2^i - \xi_2^j \d_j \xi_1^i +
	g^{ij}(\xi_1 \d_j \xi_2 - \xi_1 \d_j\xi_2). 
\end{gather}
That the algebra of the charges reproduces the algebra of the asymptotic symmetries follows in fact from general theorems \cite{Brown:1986ed}.

\subsection{Gauge transformations}

The gauge transformations are given by
\begin{equation}
    \delta_{\epsilon, \mu, \lambda} B_{ij} = \d_i\epsilon_j - \d_j
    \epsilon_i, \quad
    \delta_{\epsilon, \mu, \lambda} \Psi_i = \mu_i, \quad
    \delta_{\epsilon, \mu, \lambda} \Phi = \lambda,
\end{equation}
with 
\be
\epsilon_i = \xbar \epsilon _i + O\Big(\frac{1}{r}\Big), \quad \mu_i = \frac{\xbar \mu_i}{r} + O\big(\frac{1}{r^2}\Big) , \quad \lambda = \frac{\xbar \lambda}{r} + O\big(\frac{1}{r^2}\Big)
\ee
In polar coordinates, this yields for the leading orders
\begin{eqnarray} 
&& \delta_{\epsilon, \mu, \lambda} \xbar B_{rA} = \xbar \epsilon_A - \partial_A \xbar \epsilon_r, \qquad  \delta_{\epsilon, \mu, \lambda} \xbar B_{AB} = \d_A \xbar \epsilon_B - \d_B \xbar \epsilon_A \label{eq:VarAsF1}\\
 && \delta_{\epsilon, \mu, \lambda} \xbar \Psi_r  = \xbar \mu_r, \qquad  \delta_{\epsilon, \mu, \lambda} \xbar \Psi_A  = \xbar \mu_A, \qquad  \delta_{\epsilon, \mu, \lambda} \xbar \Phi  = \xbar \lambda \label{eq:VarAsF2}
 \end{eqnarray}
where
\begin{gather}
    \epsilon_A = r \xbar \epsilon_A + O(1), \quad 
    \epsilon_r = \xbar \epsilon_r + O(r^{-1}), \quad
    \mu_A = \xbar \mu_A + O(r^{-1}), \\
    \mu_r = r^{-1} \xbar \mu_r + O(r^{-2}),\quad
    \lambda = r^{-1} \xbar \lambda + O(r^{-2}).
\end{gather}
The associated generator is easily computed
\begin{multline}
    G_{\epsilon, \mu,\lambda} = \int d^3x \Big( \epsilon_i(-2\d_j
      \pi^{ji}) + \mu_i p^i + \lambda \pi_\Phi\Big) 
    + 2 \oint d^2x \, \xbar \pi^{rA} \xbar \epsilon_A \\
    + 2 \oint d^2x \sqrt {\xbar \gamma} \, \Big( \xbar \Psi^A (\xbar \epsilon_A -
    \d_A \xbar \epsilon_r) - \xbar \mu^A \xbar B_{rA} - \xbar \mu_r \xbar
    \Phi + \xbar \lambda \, \xbar \Psi_r\Big). \label{eq:ChargeFinal}
\end{multline}

The charges (\ref{eq:ChargeFinal}) exhibit a number of interesting features.  First we note that they are generically non-zero, since the asymptotic values of the gauge parameters are generically non-zero.  This means that they generically define improper gauge transformations \cite{Benguria:1976in}.  It is only when $ G_{\epsilon, \mu,\lambda} \approx 0$ that the transformations are proper gauge transformations with no effect on the physical system.  The charges involve asymptotic features of the $2$-form field and of the additional degrees of freedom, and differ from the conserved line integrals that measure the strength of string sources when these are present.

Second, we observe that the transformations generated by $G_{\epsilon, \mu,\lambda} $ are reducible.  If $\xbar \epsilon_A = \partial_A \xbar \chi$ and $\xbar \epsilon_r = \xbar \chi$ (for some $\chi$) i.e.,  $\xbar \epsilon_A = \partial_A \xbar \epsilon_r$, then the terms in $G_{\epsilon, \mu,\lambda}$ containing $\xbar \epsilon_i$ vanish.  This is clear for the term in $\xbar \Psi^A$.  This is also clear for the term in $\xbar \pi^{rA}$ if one recalls that $\d_A \xbar \pi^{rA} = 0$.

Third we note that one can express the contribution to $ G_{\epsilon, \mu,\lambda}$ involving $\xbar \pi^{rA}$ in terms of the original scalar field.  One has $\xbar \pi^{rA} = \varepsilon^{AB} \partial_B \xbar \phi $ and thus
\be 
2 \oint d^2x \, \xbar \pi^{rA} \xbar \epsilon_A = 2 \oint d^2x \, \varepsilon^{AB} \partial_B \xbar \phi \, \xbar  \epsilon_A.
\ee
(One can also write 
$- 2 \oint d^2x \,  \sqrt {\xbar \gamma} \, \xbar \mu^A \xbar B_{rA}$ in terms of the conjugate momentum $\pi$ but the expression is non local.)  If $\xbar \Psi^A$   happens to be even, this is the only contribution to $ G_{\epsilon, \mu,\lambda}$ involving $\xbar \epsilon_A^{odd}$.  If  $\xbar \Psi^A$ has also an odd part, $2 \oint d^2x \, \varepsilon^{AB} \partial_B \xbar \phi \, \xbar  \epsilon_A$ coincides with the generator of the $\epsilon_A$ transformation only in the ``frame'' where $\xbar \Psi^A_{odd} = 0$.  That these two quantities are not always equal is as it should since $ G_{\epsilon, \mu=0,\lambda=0}$ and $2 \oint d^2x \, \varepsilon^{AB} \partial_B \xbar \phi \, \xbar  \epsilon_A$ transform differently under improper asymptotic symmetries: $2 \oint d^2x \, \varepsilon^{AB} \partial_B \xbar \phi \, \xbar  \epsilon_A$ is invariant, while $ G_{\epsilon, \mu=0,\lambda=0}$ transforms due to the central charges that are present in the algebra.

Indeed the algebra of the asymptotic charges reproduces the algebra of the asymptotic transformations that they generate up to central charges that may be present \cite{Brown:1986ed}.  In our case, the transformations commute, so everything boils down to the central charges which turn out to be present and non-trivial given that the algebra of the transformations is abelian.

 One evaluates the central charges as follows,
 \begin{eqnarray} 
[G_{\epsilon_1, \mu_1,\lambda_1}, G_{\epsilon_2, \mu_2,\lambda_2} ]&=&
\delta_{\epsilon_2, \mu_2,\lambda_2}  G_{\epsilon_1, \mu_1,\lambda_1}  \\
&=&2 \oint d^2x \sqrt {\xbar \gamma} \, \Big( \xbar \mu^{(2)A} (\xbar \epsilon_A^{(1)} -
    \d_A \xbar \epsilon^{(1)}_r) - \xbar \mu^{(1)A}(\xbar \epsilon_A^{(2)} -
    \d_A \xbar \epsilon_r^{(2)}) \Big) \nonumber \\
    && + 2 \oint d^2x \sqrt {\xbar \gamma} \, \Big(  - \xbar \mu_r^{(1)} \xbar
    \lambda^{(2)} + \xbar \lambda^{(1)} \, \xbar \mu_r^{(2)}\Big)
\end{eqnarray}

A similar computation yields the poisson brackets between the improper
gauge symmetries and the Poincar\'e transformations
\begin{equation}
    [G_{\epsilon, \mu,\lambda},P_{\xi,\xi^i}] =
      \delta_{\xi,\xi^i}G_{\epsilon, \mu,\lambda} = 
      G_{\hat\epsilon, \hat\mu,\hat\lambda},
\end{equation}
where
\begin{gather}
    \hat\epsilon_i  = - \cL_{\xi^k} \epsilon_i- \xi \mu_i,\qquad
    \hat\lambda =  - \cL_{\xi^k} \lambda- \xi \nabla^i\mu_i, \\
    \hat\mu_i =  - \cL_{\xi^k} \mu_i-\xi \d_i \lambda - \nabla^j\Big( \xi(\d_j \epsilon_i -
    \d_i \epsilon_j)\Big).
\end{gather}
The resulting action of the Poincar\'e transformations on the asymptotic
symmetry parameters is then
\begin{gather}
    \delta_{(Y,b,T,W)} \xbar \epsilon_A = \cL_Y \xbar \epsilon_A+ b \xbar
    \mu_A,\qquad
    \delta_{(Y,b,T,W)} \xbar \epsilon_r =  \cL_Y \xbar \epsilon_r+ b \xbar \mu_r,\\
    \delta_{(Y,b,T,W)} \xbar \mu_A = \cL_Y \xbar \mu_A + b \d_A \xbar \lambda + b(\xbar
    \epsilon_A - \d_A \xbar\epsilon_r) + D^B \Big( b(\d_B \xbar \epsilon_A
    -\d_A \xbar \epsilon_B)\Big),\\
    \delta_{(Y,b,T,W)} \xbar \mu_r = \cL_Y \xbar \mu_r - b \xbar \lambda - D^A\Big(b(\xbar
    \epsilon_A - \d_A \xbar \epsilon_r)\Big),\\
    \delta_{(Y,b,T,W)} \xbar \lambda = \cL_Y \xbar \lambda + b\xbar \mu_r + b D^A \xbar \mu_A.
\end{gather}

\section{Conclusions}
\label{sec:Conclusions}

We have indicated here how one could relate some of the conserved quantities appearing in the scalar field theory to generators of asymptotic symmetries in the dual $2$-form formulation.   These charges act on variables that are not present in the original scalar field formulation and this explains why they cannot be viewed as generators there.  Going to the original formulation in terms of the scalar field  involves in fact a truncation.

Two major features characterize the construction: asymptotic conditions fulfilling parity conditions that involve a twist given by an improper gauge transformations and  introduction of extra surface degrees of freedom at spatial infinity.  While the first step is common to electromagnetism and gravity, the second one was found so far to be needed for the same reason (necessity to make the boosts canonical transformations) only in electromagnetism \cite{Henneaux:2018hdj}.  Perhaps the description of superrotations \cite{Banks:2003vp,Barnich:2009se}, which has not yet been achieved in the canonical formalism on spacelike Cauchy surfaces, would also need the introduction of surface degrees of freedom.

Even though the followed procedures are rather different, it is reassuring that some of the features encountered in our analysis at spatial infinity are also present at null infinity \cite{Campiglia:2018see}: need to introduce new surface degrees of freedom, absence of generator associated with the zero mode of the scalar field $\phi$, no action of the symmetry transformations on  $\phi$.  The detailed connection between the two approaches has not been worked out, however, and would deserve further study.

The classical vacuum of the theory is naturally defined by setting all fields equal to zero. The infinite-dimensional symmetry exhibited here acts non trivially on this vacuum. As it follows from (\ref{eq:VarAsF1}) and (\ref{eq:VarAsF2}),  improper gauge transformations generate non vanishing values of the asymptotic fields so that the orbit of the vacuum is non trivial.  In other words, the vacuum is  degenerate, as in the electromagnetic or gravitational cases.  The energy of all these vacua is the same and equal to zero.

We have also studied in detail the behaviour of the scalar field as one approaches null infinity.  This constitutes an excellent laboratory for exhibiting in a simple context free from gauge invariance questions the type of non-analytic behaviour that emerges as one takes that limit.

Finally, we have shown that the algebra of the canonical generators of asymptotic symmetries is a non trivial central extension of the abelian algebra of the symmetries.  As a result, contrary to the gravity or electromagnetic cases, the various vacua are characterized by different values of the charges asociated with the new symmetry.


\section*{Acknowledgments} 
This work was partially supported by the ERC Advanced Grant ``High-Spin-Grav'' and by FNRS-Belgium (convention IISN 4.4503.15).

\begin{appendix}

\section{Ultraspherical polynomials and functions of the second kind}
\label{app:ultraspherical}

The relevant equation for our analysis is
\begin{gather}
	\label{eq:alternJacobieq}
    (1-s^2) \d_s^2 Y^{(\lambda)}_n + (2\lambda -3) s \d_s Y^{(\lambda)}_n + (n+1)(n+2\lambda -1)
    Y^{(\lambda)}_n =
	0,\\
    \lambda = k + \frac {d-3}{2}, \quad n = l-k.
\end{gather}
($d=4$ in our case so that $\lambda = k + \frac12$, but the analysis proceeds in the same way for arbitrary spacetime dimension $d$.  Note that $2 \lambda$ is always an integer.)
\begin{itemize}
    \item $n\ge 0$,
We can easily obtain the solutions for $n\ge 0$ by considering the following
rescaling
\begin{equation}
    Y^{(\lambda)}_n(s)  = (1-s^2)^{\lambda-\frac 1 2 } \psi^{(\lambda)}_n(s).
\end{equation}
The equation for $\psi_n^{(\lambda)}$ then takes the form
\begin{equation}
	\label{eq:diffJacobi}
	(1-s^2) \d_s^2 \psi_n  - (2\lambda +1 ) s \d_s
	\psi_n + n (n+2 \lambda) \psi_n = 0.
\end{equation}
For $\lambda > \frac 1 2 $ and $n=\NN$, this equation has a polynomial
solution known as ultraspherical polynomial or
Gegenbauer's polynomial $P^{(\lambda)}_n$ \cite{Ultra}.
These polynomials satisfy
\begin{equation}
	P_n^{(\lambda)}(-s) = (-)^n P_n^{(\lambda)}(s), \qquad P^{(\lambda)}_n(1) = \left
	(\begin{array}{c}n+2\lambda -1 \\ n \end{array}
	\right).
\end{equation}
and can be constructed using the following recurrence
formula
\begin{gather}
	nP^{(\lambda)}_{n}(s) = 2 (n+\lambda-1)s P^{(\lambda)}_{n-1}(s) -
	(n+2\lambda -2) P^{(\lambda)}_{n-2}(s), \qquad n>1,\\
	P^{(\lambda)}_0(s) = 1, \qquad P^{(\lambda)}_1(s) = 2\lambda s.
\end{gather}
When $\lambda = \frac 1 2$, we recover Legendre Polynomials. 
The function of the second kind $Q^{(\lambda)}_n$ is the solution of the
differential equation \eqref{eq:diffJacobi} which is linearly independent of
$P^{(\lambda)}_n$. The full set can be constructed using the same recurrence
relation with a different starting point:
\begin{gather}
	nQ^{(\lambda)}_{n}(s) = 2 (n+\lambda-1)s Q^{(\lambda)}_{n-1}(s) -
	(n+2\lambda -2) Q^{(\lambda)}_{n-2}(s), \qquad n>1,\\
	Q^{(\lambda)}_0(s) = \int_0^s(1-x^2)^{-\lambda-\frac 1 2} dx, 
	\qquad Q^{(\lambda)}_1(s) = 2 \lambda s Q^{(\lambda)}_0(s) -
	(1-s^2)^{-\lambda + \frac 1 2 }.
\end{gather}
They take the general form
\begin{equation}
	Q^{(\lambda)}_n(s) = P^{(\lambda)}_n(s) Q^{(\lambda)}_0(s) + 
	R^{(\lambda)}_n(s)(1-s^2)^{-\lambda + \frac 1 2 },
\end{equation}
where $R^{(\lambda)}_n$ are polynomials of degree $n-1$ and satisfy
$Q^{(\lambda)}_n(-s) = (-)^{n+1} Q_n^{(\lambda)}(s)$.
The values of $\lambda$ relevant for our analysis are half integers and
integers. In these cases $Q^{(\lambda)}_n(s)$ diverges at $s=\pm 1$:
for $\lambda = \frac 1 2 $, the Legendre function of the second kind diverges logarithmically while the other values of
$\lambda$ lead to
\begin{equation}
	\lim_{s\to 1} (1-s^2)^{\lambda - \frac 1 2}Q_n^{(\lambda)}(s) = \frac
	1{2\lambda - 1}, \qquad
	k=2,3,\ldots.
\end{equation}

The general solution for $Y^{(\lambda)}_n$ is then given by
\begin{gather}
    Y^{(\lambda)}_n(s) =
	A \tilde P^{(\lambda)}_n(s) + B
    \tilde Q^{(\lambda)}_n(s), \\
\tilde P^{(\lambda)}_n(s) = (1-s^2)^{\lambda-\frac 1 2
    }P^{(\lambda)}_n(s), \qquad \tilde Q^{(\lambda)}_n(s) =
    (1-s^2)^{\lambda-\frac 1 2 }Q^{(\lambda)}_n(s), \quad \forall n\ge 0,\\
    \tilde P^{(\lambda)}_n(-s) = (-)^n \tilde P^{(\lambda)}_n(s), \qquad
    \tilde Q^{(\lambda)}_n(-s) = (-)^{n+1} \tilde Q^{(\lambda)}_n(s).
\end{gather}
For all values of $\lambda$, the $\tilde Q$ branch of $Y$ will dominate in the limit
$s\to\pm 1$. If $\lambda = \frac 1 2$, the $\tilde Q$ branch will diverge
logarithmically while the $\tilde P$ branch will be finite. For all half integer
value of $\lambda \ge 1$, the $\tilde Q$ branch will be finite and will tend to a
non-zero constant at $s=\pm 1$ while the $\tilde P$ branch will go to zero.

\item $0>n>1-2\lambda$. The pattern here is similar: we have a polynomial
branch and a "second" class branch. The two sets of solutions can be
constructed with the following recurrence relation:
\begin{gather}
    (2\lambda + n -1) \tilde P^{(\lambda)}_{n-1}(s) = 2 (n+\lambda) s
    \tilde P^{(\lambda)}_n(s) - (n+1)
    \tilde P^{(\lambda)}_{n+1}(s),\qquad \lambda>2, \quad n < -2, \\
    \tilde P^{(\lambda)}_{-1}(s) =\int_0^s (1-x^2)^{\lambda -
    \frac 3 2} dx, \quad \tilde P^{(\lambda)}_{-2}(s) = s \tilde
    P^{(\lambda)}_{-1}(s) +  \frac 1 {2 (\lambda -1)} (1-s^2)^{\lambda - \frac 1 2 },
\end{gather}
the polynomial branch being given by the following starting point
\begin{equation}
    \tilde Q^{(\lambda)}_{-1}(s) = 1, \quad \tilde Q^{(\lambda)}_{-2}(s) =
    s.
\end{equation}
In order to prove this recurrence, one needs the following relation
\begin{equation}
    (1-s^2) \d_s \tilde P^{(\lambda)}_n = (n+1) (s \tilde P^{(\lambda)}_n
    -\tilde P^{(\lambda)}_{n+1})
\end{equation}
and its equivalent in terms of $\tilde Q$. We have chosen the notation in
order to have consistent parity conditions 
\begin{equation}
    \tilde P^{(\lambda)}_n(-s) = (-)^n \tilde P^{(\lambda)}_n(s), \qquad
    \tilde Q^{(\lambda)}_n(-s) = (-)^{n+1} \tilde Q^{(\lambda)}_n(s).
\end{equation}

The general solution for $Y^{(\lambda)}_n$ keeps the form
\begin{equation}
    Y^{(\lambda)}_n(s) =
	A \tilde P^{(\lambda)}_n(s) + B
    \tilde Q^{(\lambda)}_n(s), \qquad  0>n>1-2\lambda.
\end{equation}
An important difference from the regime $n\ge 0$ is that both branches
have now the same asymptotic behaviour in the limit $s\to \pm 1$: they
both tend to a non-zero finite value. In particular, if $\lambda$ is a half integer,
both branches are polynomials.
\end{itemize}
Except for the already solved case where $l=k=0$ and $d=4$, $n+2\lambda = l +
k +d-4>0$ which means that we don't have to consider the solutions in the
regime $n\le - 2\lambda$.

\section{Symplectic Form - Alternative approaches - Improper gauge fixings}
\label{app:Minimal}

\subsection{Relation between the kinetic terms of the scalar field action and the $2$-form action}
Under the change of variables (\ref{eq:CofV}), the kinetic terms of the scalar field action (\ref{eq:Ham0}) and of the $2$-form action (\ref{eq:Ham2}) can be checked to be related as
\begin{equation}
    \int d^3x \pi \d_t \phi = d_t \int d^3x \frac {1}{\sqrt 2}
    \phi \epsilon^{ijk} \d_i B_{jk} + \int d^3x \, \pi^{ij} \d_t B_{ij} + \oint
    d^2x\, \xbar \pi^{AB} \d_t \xbar B_{AB}.
\end{equation}
 The total derivative term $d_t \int d^3x \frac {1}{\sqrt 2}
    \phi \epsilon^{ijk} \d_i B_{jk}$ corresponds to a change of representation, from the ``coordinate representation'' to the ``momentum representation'' (keeping $B_{ij}$ fixed at the time boundaries amounts to keeping the momentum $\pi$ fixed there) and has an analog for systems with a finite number of degrees of freedom.  More unusual is the surface term at spatial infinity $\oint
    d^2x\, \xbar \pi^{AB} \d_t \xbar B_{AB}$, which modifies the symplectic structure of the $2$-form.  
    
    The Hamiltonians are  easily verified to be the same.

\subsection{Description of minimal approach}

Since the scalar field system presents no problem with boosts, one expects that keeping the surface term $\oint
    d^2x\, \xbar \pi^{AB} \d_t \xbar B_{AB}$ in the $2$-form kinetic term should lead to a formulation that is relativistically invariant without extra degrees of freedom.  This is indeed the case and constitutes the ``minimal solution'' to the difficulty with boosts pointed out in the text.
    
With the action
\begin{eqnarray}
    S[B_{ij}, \pi^{ij}; B_{0}] &=& \int d^4x \left\{ \pi^{ij} \d_t B_{ij} -
    B_{0i}(-2 \d_j \pi^{ji}) - \Big( \frac 1 {2\sqrt g} \pi^{ij}\pi_{ij} +
    \frac {\sqrt g}{6} H_{ijk}H^{ijk}\Big)\right\}\nonumber \\
    &&+ \int dt  \oint
    d^2x\, \xbar \pi^{AB} \d_t \xbar B_{AB},
\end{eqnarray}
leading to the symplectic form 
\be
\Omega^{min} =  \int d^3 x \, d_V \pi^{ij} \wedge d_V B_{ij} + \oint
    d^2x\, d_V \xbar \pi^{AB} \wedge d_V \xbar B_{AB}, 
\ee
the Poincar\'e transformations are all canonical transformations.  
Their generators are given by
\begin{flalign}
    \label{eq:PoincMinI}
    P_{(\xi,0)} & = \int d^3x\, \xi \Big( \frac 1 {2\sqrt g} \pi^{ij}\pi_{ij} +
    \frac {\sqrt g}{6} H_{ijk}H^{ijk}\Big),\\
    \label{eq:PoincMinII}
    P_{(0,\xi^i)} & = \int d^3x \, \xi^i H_{ijk} \pi^{jk}.
\end{flalign}

Gauge transformations are generated by parameters of the form
\begin{equation}
    \epsilon_A = r \xbar \epsilon_A + O(1), \qquad \epsilon_r = \xbar
    \epsilon_r + O(r^{-1}).
\end{equation}
The associated generator is easily computed
\begin{equation}
    G_\epsilon = \int d^3x \epsilon_i \mathcal G^i + 2 \oint d^2x \xbar
    \epsilon_A (\xbar \pi^{rA} - \d_B \xbar \pi^{BA}) 
    = \int d^3x \epsilon_i \mathcal G^i,
\end{equation}
where we used the fact that we imposed the constraints asymptotically.   This has the effect of setting all charges on-shell to zero:  all
allowed gauge transformations have well-defined generators that are
proportional to the constraints. In the minimal formulation, there are therefore no
improper gauge transformations left and no surface charges.

\subsection{Hyperbolic coordinate formulation}

The same conclusion follows from an analysis in hyperbolic coordinates, which we briefly discuss. 

The asymptotic conditions on $B_{\mu\nu}$ in hyperbolic coordinates  that are compatible with the
ones given previously for $\phi$ are
\begin{gather}
    B_{\eta a} = \xbar B_{\eta a} + O(\eta^{-1}), \qquad 
    B_{ab} = \eta \xbar B_{ab} +O(1)\\
    C_{\eta a b} = \xbar B_{ab} + \d_b \xbar
    B_{\eta a} - \d_a \xbar B_{\eta b} + O(\eta^{-1}).
\end{gather}
(modulo parity conditions that play no role in the discussion of this subsection).

The bulk action $ S_{bulk}[B_{\mu\nu}] = \int d^4x \, \frac {\sqrt{-g}}{6} C_{\mu\nu\rho}
        C^{\mu\nu\rho}$ does not lead to a  well defined variaional principle with these boundary conditions. Evaluating the
variation of the action, we get
\begin{equation}
    \delta S_{bulk} = \int d^4x \, \d_\rho \left(\sqrt {-g} C^{\rho
    \mu\nu}\right) \delta B_{\mu\nu}
    + \oint d^3x \, \sqrt{-h} h^{ac} h^{bd} \left(\xbar B_{ab} + \d_b \xbar
    B_{\eta a} - \d_a \xbar B_{\eta b}\right) \delta B_{cd}
\end{equation}
which is not zero even on-shell due to the surface term.  Given that time translations in hyperbolic coordinates involve boosts, this is the way the difficulty with boosts appears in the hyperbolic formulation. 

The minimal way to
take care of the surface term is to modify the action by a boundary term
\begin{equation}
    S = \int d^4x \, \frac {\sqrt{-g}}{6} C_{\mu\nu\rho}
        C^{\mu\nu\rho} 
    - \oint d^3x \, \sqrt{-h} \frac 1 2 \xbar C_{\eta a b} \xbar {C_{\eta}}^{a b},
      \end{equation}
with 
\be
\xbar C_{\eta a b}= \xbar B_{ab}
      + \d_b \xbar B_{\eta a} - \d_a \xbar B_{\eta b}
      \ee
The variation of the modified action leads  to
\begin{equation}
    \delta S = \int d^4x \, \d_\rho \left(\sqrt {-g} C^{\rho
    \mu\nu}\right) \delta B_{\mu\nu}
    + 2\oint d^3x \, \sqrt{-h} h^{ac}  \mathcal D^b\left(\xbar B_{ab} + \d_b \xbar
    B_{\eta a} - \d_a \xbar B_{\eta b}\right) \delta
    B_{\eta c}.
\end{equation}
The extra boundary term  is now proportional to a bulk equation of motion:
\begin{equation}
    \d_\rho(\sqrt {-g} C^{\rho \eta a}) = \eta^{-3}\sqrt{-h} h^{ac} \mathcal D^b\left(\xbar B_{ab} + \d_b \xbar
    B_{\eta a} - \d_a \xbar B_{\eta b}\right) + O(\eta^{-4})
\end{equation}
so that the action is truly stationary when the bulk EOM hold.

Due to the fact that both the bulk and the boundary term in the action are
build out of gauge invariant objects, there are no extra constraints on
the gauge parameters. All parameters of the form
\begin{equation}
    \epsilon_\eta = \xbar \epsilon(s,x^A) + O(\eta^{-1}), \qquad
    \epsilon_a = \eta \xbar \epsilon_a(s, x^A) + O(1),
\end{equation}
generate a symmetry of the action. The absence of extra constraints on the
boundary values of these parameters, in particular the possibility of an
arbitrary dependence on time, means that they are proper gauge
transformations of the system. This theory does not have any non-trivial
boundary charges, as we saw previously in the standard Hamiltnian formulation.

\subsection{Minimal formulation as resulting from an improper gauge fixing}

We can view the minimal formulation as resulting from an improper gauge fixing of the full theory with action (\ref{eq:FullAction2B}).   From that point of view, the minimal formulation is thus a truncation of the richer theory that exhibits all the symmetries.

Using the (improper) gauge transformations with parameter $\mu_i$, one can impose the following conditions
    \be
    \xbar \Psi_r = 0 , \quad \xbar \pi^{rA} + \sqrt{\xbar \gamma} \, \xbar \Psi^A = 0. \label{eq:ImpGF}
    \ee
These transformations are improper and set to zero the conserved charges associated with  $\xbar\epsilon_A $ and $\xbar\lambda$. These conditions eliminate therefore
all asymptotic symmetries: $\xbar\mu_i$ is frozen by the (improper) gauge conditions while the other transformations become pure gauge since they have a vanishing charge for all remaining configurations. 

With the supplementary conditions (\ref{eq:ImpGF}), the boundary term in the symplectic structure reduces to
  \be  -2 \oint d^2x \sqrt\gamma d_V \xbar\pi^{rA} d_V \xbar B_{rA}. \label{eq:SympFB17}
  \ee
Using the asymptotic behaviour of the fields, including the parity conditions and the asymptotic form of the constraints, it is then easy to verify that (\ref{eq:SympFB17}) becomes
\be    + \oint d^2x \sqrt\gamma d_V \xbar\pi^{AB} d_V \xbar B_{AB},
\ee
which is precisely the boundary term of the minimal formulation.

The original Poincar\'e transformations given in
\eqref{eq:fullponcareI}-\eqref{eq:fullponcareEnd} do not preserve the
improper gauge fixing. Nevertheless, one can check that the compensating
improper gauge transformation needed to restore the gauge fixing
conditions \eqref{eq:ImpGF} is integrable and that the resulting total generator
match the one given in \eqref{eq:PoincMinI} and \eqref{eq:PoincMinII} for the Poincar\'e trasformatinos
of the minimal description.

We should close this section by stressing that the process of gauge fixing improper gauge transformations is ``illegal'' since these do change the physical state of the system \cite{Benguria:1976in}.  The passage  to the minimal formulation truncates physical (surface) degrees of freedom from the point of view of the complete theory.  As a result, one looses in particular the symmetries that act on these  degrees of freedom that have been improperly dropped.

\section{Tensors, pseudo-tensors and parity conditions}
\label{app:Pseudo}

A function $f(x^i)$ is even or odd according to whether $f(-x^i) = f(x^i)$ or $- f(x^i)$.  We have contemplated tensor fields of various types and considered the parity of their components viewed as functions.  A more intrinsic definition can be given. 

Let $T$ be a tensor field.  Let $T^*$ be its image under the central reflection (antipodal map)   $y^i \rightarrow y'^i = f^i(y^m)$ (which is $x^i \rightarrow - x^i$ in cartesian coordinates, but we formulate the definition in arbitrary coordinates).    For instance, one has for a covariant tensor of rank $k$,
\be
{T^*}_{i_1 \cdots i_k} (y^m)= \frac{\partial f^{j_1}}{\partial y^{i_1}} \cdots \frac{\partial f^{j_k}}{\partial y^{i_k}}T_{j_1 \cdots j_k} (f^m(y^n)) \label{eq:TransfRefl}
\ee 
A tensor field will have definite parity if $T^* = \pm T$. The $+$ case defines even tensors, the $-$ case defines odd tensors.

In cartesian coordinates $x'^i = - x^i$ and the transformed tensor $T^*$  is\footnote{We consider only covariant tensors (``with all indices down''), as one can do by using the cartesian metric, which is even.},
\be  {T^*}_{i_1 \cdots i_k} (x^m)= \frac{\partial x'^{j_1}}{\partial x^{i_1}} \cdots \frac{\partial x'^{j_k}}{\partial x^{i_k}}T_{j_1 \cdots j_k} (x'^m)= (-1)^k T_{i_1 \cdots i_k}(- x^m). \label{eq:TransfCart}
\ee 
From (\ref{eq:TransfCart}), one sees that the components in cartesian coordinates of an even tensor of even (respectively, odd) rank are even (respectively, odd), while the components in cartesian coordinates of an odd tensor of even (respectively, odd) rank are odd (respectively, even).

The parity properties of the components depend on the coordinate system.  Two other coordinate systems have been considered in the text:
\begin{itemize}
\item (``Unorthodox'') polar coordinates $(r, x^A)$, where $r = \sqrt{\sum_i \left(x^i\right)^2}$ and $x^A$ are coordinates on the $2$-sphere such that the central reflection reads
\be
r \rightarrow r' = r, \qquad x^A \rightarrow x'^A = - x^A
\ee
For instance, $(x^A) = (\frac{x^1}{r}, \frac{x^2}{r})$.
\item Standard polar coordinates $(r, \theta, \varphi)$ for which the central reflection reads
\be
r \rightarrow r' = r, \qquad \theta \rightarrow \theta' = \pi - \theta, \qquad \varphi' = \varphi + \pi.
\ee
\end{itemize}
Note that $dx^i \rightarrow - dx^i$, $dr \rightarrow dr$, $dx^A \rightarrow - dx^A$, $d \theta \rightarrow - d \theta$ and $d \varphi \rightarrow  d \varphi$.
Note also for later purposes that while the Jacobian $J$ of the central reflexion is equal to $-1$ in Cartesian coordinates, it is equal to $+1$ in the polar coordinates $(r, x^A)$ and to $-1$ in the polar coordinates  $(r, \theta, \varphi)$.

To illustrate how the components of a tensor of definite parity behave under  reflection, consider for instance an even $2$-form $C \equiv \frac 12 C_{ij} dx^i \wedge dx^j$. The condition $C^* = C$, equivalent to even $C_{ij}$'s in cartesian coordinates,  implies that the $C_{rA}$ are odd and the component $C_{AB}$ even.  By contrast, $C_{r\theta}$ and $C_{\theta \varphi}$ are odd and $C_{r \varphi}$ is even. 

Tensor densities involving $\sqrt{g}$ (like the conjugate momenta) behave under inversion as tensors in the coordinate systems considered here since $\sqrt{g'} = \sqrt{g}$ in all cases.  This is not true
for pseudo-tensors, which involve the determinant $J$ of the reflection (with its sign) in their transformation law. For instance, in cartesian coordinates
\be  {T^*}_{i_1 \cdots i_k} (x^m)= J \, \frac{\partial x'^{j_1}}{\partial x^{i_1}} \cdots \frac{\partial x'^{j_k}}{\partial x^{i_k}}T_{j_1 \cdots j_k} (x'^m)= (-1)^{k+1} T_{i_1 \cdots i_k}( - x^m) \label{eq:TransfCartPseudo}
\ee 
with 
\be
J = \det \left(\frac{\partial x'^m} {\partial x^n} \right)
\ee
If we keep the convention that an even pseudo-tensor must be such that $T^* = T$, and that an odd pseudo-tensor fulfills $T^* = - T$, as it is natural from the intrinsic point of view, one then finds that the parity of the components are reversed with respect to what they are in the tensor case in coordinate systems where $J = -1$, but are the same in coordinate systems for which $J = 1$.  

To take the example of an odd pseudo-$2$-form $B$ ($B^* = -B$) relevant to our discussion in the text, one finds that
\begin{itemize}
\item the cartesian components $B_{ij}$ are even ($J=-1$);
\item $B_{r A}$ are even and $B_{AB}$ is odd ($J=1$);
\item $B_{r \theta}$ and $B_{\theta \varphi}$ are odd while $B_{r \varphi}$ is even ($J=-1$).
\end{itemize}
The conjugate momentum is an even pseudo-bivector (density)$\pi$ ($\pi^* = \pi$) and so
\begin{itemize}
\item the cartesian components $\pi^{ij}$ are odd ($J=-1$);
\item $\pi^{r A}$ are odd and $\pi^{AB}$ is even ($J=1$);
\item $\pi^{r \theta}$ and $\pi^{\theta \varphi}$ are even while $\pi^{r \varphi}$ is odd ($J=-1$).
\end{itemize}

It is instructive to check compatibility of this behaviour with the relationship $\pi^{ij} = \epsilon^{ijk} \partial_k \phi$ with an even scalar field $\phi$.  One finds:
\begin{itemize}
\item The cartesian components $\pi^{ij}$ are indeed odd since $\partial_k \phi$ is odd.
\item The $(r,A)$ components fulfill: the components $\pi^{r A} = \epsilon^{AB} \partial_B \phi$ are odd since $\partial_B \phi$ is odd; the component  $\pi^{AB} = \partial_r \phi $ is even since $\partial_r \phi$ is even.
\item The components in standard polar coordinates fulfill: the components $\pi^{r \theta}= \partial_\varphi \phi$ and $\pi^{\theta \varphi} = \partial_r \phi$ are even while $\pi^{r \varphi} = - \partial_\theta \phi$ is odd. This matches the parity of the components of the gradients of $\phi$.
\end{itemize}
Take for instance $\phi = \sin \theta$.  The only non-vanishing component of the conjugate momentum in standard polar coordinates is $\pi^{r \varphi}$, which is equal to $\pi^{r \varphi} = - \cos \theta$, which is odd.  In the coordinates $(r, x^A)$, with $x^1 = \sin \theta \cos \varphi$ and $x^2 = \sin \theta \sin \varphi$, one gets $ \pi^{r1} = D \frac{\partial r}{\partial y^m} \frac{\partial x^1}{\partial y^n} \pi^{mn} = \sin \varphi$ and $\pi^{r2}= - \cos \varphi $ which are both odd.  Here $D$ is the determinant of the Jacobian matrix of the transformation, $D= \left(\sin \theta \cos \theta \right)^{-1}$.

\end{appendix}

\end{document}